\documentclass[10pt,journal]{IEEEtran}
\IEEEoverridecommandlockouts
\usepackage{stfloats}
\usepackage{graphicx}
\usepackage{amsmath,amsthm,amsfonts,amssymb}
\usepackage{cite,hyperref}
\usepackage{bm}
\usepackage{bbm}
\usepackage{url}
\usepackage{array}
\usepackage{color,soul}
\usepackage{multirow}
\usepackage{booktabs}
\usepackage[table,xcdraw]{xcolor}
\usepackage{enumitem}
\usepackage{subcaption}

\usepackage[english]{babel}
\usepackage{algorithm}
\usepackage{algorithmic}
\usepackage{setspace}
\usepackage[english]{babel}

\usepackage{stfloats}
\newcounter{CWtempeq}
\usepackage{refcount}

\theoremstyle{plain}
\newtheorem{thm}{Theorem}
\newtheorem{lem}[thm]{Lemma}

\newtheorem{prop}[thm]{Proposition}
\newtheorem{cor}[thm]{Corollary}

\allowdisplaybreaks[4]

\begin{document}
\title{Harnessing Inter-Waveguide Evanescent Coupling in Pinching-Antenna Systems}

\author{
Dengke Wei,
Runxin Zhang,
Yulin Shao,
Shaodan Ma

\thanks{D. Wei and S. Ma are with the State Key Laboratory of Internet of Things for Smart City, University of Macau, Macau, China (dengke.wei@connect.um.edu.mo, shaodanma@um.edu.mo).}
\thanks{R. Zhang is with the Department of Electronic Engineering, Tsinghua University, Beijing, China (rxz@mail.tsinghua.edu.cn)
}
\thanks{Y. Shao is with the Department of Electrical and Electronic Engineering, The University of Hong Kong, Hong Kong, China (ylshao@hku.hk). 
}
}

\maketitle

\begin{abstract}
In multi-waveguide pinching-antenna systems (PASS), waveguides are typically modeled as independent signal-transport paths. However, when dielectric waveguides are closely spaced, their evanescent electromagnetic fields overlap, allowing guided signals to exchange energy before radiation. While such inter-waveguide coupling is conventionally regarded as undesirable crosstalk, this paper reveals an opportunity to harness it as an additional analog beamforming resource. The key insight is that coupling, through its interaction with sequential signal extraction by distributed pinching antennas (PAs), can reshape coherent signal combining at the receivers without increasing the available signal power. Consequently, stronger coupling does not necessarily yield better performance; rather, the coupling-induced phase relationships determine whether the radiated signals combine constructively or destructively. To characterize and exploit this mechanism, we establish a physics-based end-to-end channel model, derive coupling-aware waveguide-spacing principles and achievable-rate bounds, and develop joint waveguide deployment, PA placement, and digital precoding algorithms for multi-hotspot communications. Numerical results demonstrate substantial sum-rate improvements over conventional fixed-deployment and optimized minimum-spacing PASS benchmarks.
These findings establish inter-waveguide evanescent coupling as a new physical degree of freedom for PASS channel reconfiguration.
\end{abstract}


\section{Introduction}
Pinching-antenna systems (PASS) have recently attracted considerable attention as a promising architecture for reconfigurable wireless communications \cite{ding2025flexible,xu2026generalized}. By transporting radio-frequency (RF) signals through low-loss dielectric waveguides and extracting them into free space using pinching antennas (PAs), PASS enables flexible radiation at locations close to intended users \cite{liu2026pinching}. Unlike conventional antenna arrays with fixed radiating elements, PASS can adapt its radiation geometry to changing user distributions, thereby reducing free-space propagation losses and improving communication performance \cite{suzuki2022pinching,wang2025modeling,zhang2026directional,wei2026multi}. Such flexibility is particularly appealing in multi-hotspot scenarios \cite{xu2026environmentaware,ding2026movable}, where communication demands are concentrated in localized regions. To further enhance spatial flexibility and multi-user transmission capabilities, recent studies have extended PASS from single-waveguide configurations to multi-waveguide architectures, where multiple RF-fed waveguides and their associated PAs can be jointly configured to serve distributed users \cite{11165763,11212813}.

A growing body of research has investigated how to exploit the spatial reconfigurability of PASS. Early studies focused on PA placement \cite{chen2025dynamic,xie2025low}, signal extraction \cite{wang2025modeling,11263923}, and beamforming design \cite{11371592,wang2025modeling}, demonstrating substantial performance benefits through flexible radiation positioning. Subsequent works extended these investigations to multi-waveguide systems, addressing antenna activation \cite{11165763}, hybrid beamforming \cite{chen2026hybrid,11223640,zhang2024polarization}, interference management \cite{11223640}, and achievable-rate characterization \cite{11358830,tyrovolas2025performance}. More recently, movable-waveguide PASS has been proposed to further adapt the waveguide arrangement itself to the spatial distribution of users \cite{ding2026movable}. Despite these advances, existing multi-waveguide PASS designs predominantly rely on an important simplifying assumption: different waveguides are treated as independent signal-transport channels. Consequently, waveguide positioning is primarily exploited to optimize the geometry of free-space radiation, whereas the electromagnetic interactions among the waveguides themselves are typically neglected.

This independent-waveguide assumption is justified when neighboring waveguides are sufficiently far apart. However, it does not generally hold for closely spaced dielectric waveguides. Unlike idealized transmission lines with perfectly confined electromagnetic fields, dielectric waveguides support evanescent fields \cite{yeh2008essence,marcatili1969dielectric} that extend beyond their cores and decay exponentially into the surrounding medium. When adjacent waveguides are brought into close proximity, these fields overlap and induce coherent energy exchange between the guided modes, a phenomenon known as inter-waveguide evanescent coupling \cite{marcatili1969dielectric,marcuse1971coupling}. Such coupling is well understood in electromagnetic waveguide theory, where it is often treated as crosstalk to be suppressed when independent transmission paths are desired \cite{song2015high,mrejen2015adiabatic}. Accordingly, maintaining sufficient waveguide separation provides a convenient way to preserve the independent-waveguide model commonly adopted in PASS. Yet, as waveguide positions become increasingly configurable, enforcing large separations also rules out a potentially useful class of deployment configurations. This motivates a fundamental question: \emph{Must inter-waveguide evanescent coupling be avoided in PASS, or can it be deliberately exploited to improve wireless communication performance?}

Interestingly, the distinctive signal-radiation mechanism of PASS creates an opportunity to exploit such coupling in a way that differs from conventional coupled-waveguide systems. When multiple waveguides are electromagnetically coupled, signals launched into individual waveguides undergo coherent redistribution during propagation, leading to coupling-dependent variations in their amplitudes and phases \cite{marcuse1971coupling,christodoulides2003discretizing}. In PASS, however, these signals are not simply collected at the ends of the waveguides. Instead, multiple PAs sequentially extract portions of the guided signals at different propagation positions \cite{wang2025modeling,11263923}, while the remaining signals continue propagating and interacting with neighboring waveguides. As a result, signals radiated from different PA locations experience different coupling-induced evolutions before reaching the intended receivers. Their subsequent superposition in free space can therefore be either constructive or destructive, depending on the accumulated coupling and propagation phases. This suggests that inter-waveguide coupling, rather than merely introducing unwanted signal leakage, can be harnessed as an additional analog beamforming resource to reshape the effective wireless channels.

However, exploiting this opportunity is far from straightforward. Unlike an active amplification mechanism, evanescent coupling does not create additional signal power; it redistributes existing guided signals. Therefore, stronger coupling does not necessarily improve the received signal strength or achievable rate. The potential benefit instead lies in how coupling modifies the coherent combining of signals radiated from different PA extraction sections. This introduces a new dimension to PASS design: \emph{waveguide spacing not only determines the physical distribution of radiating elements but also controls the coherent evolution of guided signals before radiation}. To translate this physical insight into communication gains, several challenges must be addressed. 
\begin{itemize}[leftmargin=0.45cm]
    \item First, a physically grounded channel model is needed to capture the interaction between inter-waveguide coupling, guided propagation, and sequential PA extraction.
    \item Second, the relationship between coupling and communication performance must be established to identify when coupling is beneficial and how waveguide spacing should be designed.
    \item Third, in multi-user scenarios, coupling-aware waveguide deployment must be jointly coordinated with PA placement and digital precoding, since these design variables simultaneously affect coherent signal combining, radiation geometry, and inter-user interference.
\end{itemize}  

In this paper, we address these challenges by developing a comprehensive electromagnetic and communication-theoretic framework for coupled-waveguide pinching-antenna systems (CWPASS). By explicitly incorporating inter-waveguide evanescent coupling into the signal propagation and radiation processes, we establish a direct connection between waveguide electromagnetic interactions and end-to-end wireless communication performance. More importantly, we reveal that the communication benefit of coupling arises from its ability to reshape coherent combining across distributed PA extraction sections, rather than from increasing the available guided signal power. This insight enables the development of coupling-aware waveguide-spacing principles and analytical performance bounds. Building on these results, we further investigate how inter-waveguide coupling can be jointly exploited with spatial reconfiguration and digital beamforming to improve multi-hotspot communication performance.

Our main contributions are summarized as follows.
\begin{enumerate}[leftmargin=0.45cm]
    \item We introduce CWPASS, which harnesses inter-waveguide evanescent coupling as an additional physical degree of freedom for wireless channel reconfiguration. Unlike conventional multi-waveguide PASS, where individual waveguides are modeled as independent transmission paths, CWPASS exploits coherent energy exchange among neighboring waveguides to manipulate the signals radiated by distributed PAs. The key insight is that the interaction between coupled propagation and sequential PA extraction enables coupling-dependent coherent combining, providing a new mechanism for analog beamforming.
    \item Starting from the guided electromagnetic modes of rectangular dielectric waveguides, we characterize evanescent-field penetration and the resulting induced polarization currents, from which analytical inter-waveguide coupling coefficients are derived. We then establish a coupled-propagation model and explicitly incorporate sequential PA extraction and free-space radiation to obtain the end-to-end CWPASS channel. The resulting framework captures the coupling-dependent amplitude and phase evolution of signals radiated from different PA locations.
    \item We investigate how inter-waveguide coupling affects coherent signal combining across PA extraction sections. By decomposing the effective channel gain into power accumulation and inter-section coherent contributions, we identify the mechanism through which coupling influences communication performance. Based on this characterization, we derive closed-form phase-matching-based waveguide-spacing candidates for the two-waveguide case, establish coupling-aware channel-gain and achievable-rate upper bounds, and characterize the rate-ceiling scaling with the number of waveguides and necessary conditions for individual rate guarantees. These results demonstrate why coupling strength alone is insufficient to determine communication performance and provide analytical guidance for coupling-aware deployment.
    \item For multi-hotspot communications, we formulate a sum-rate maximization problem that jointly optimizes waveguide positions, PA locations, and digital precoding under transmit-power, spatial deployment, and individual rate constraints. To address the resulting nonconvex problem, we develop two complementary algorithms based on weighted minimum mean-square error and successive convex approximation (WMMSE-SCA), and signal-domain penalty dual decomposition (PDD), respectively. Numerical results demonstrate that, at 40 dBm, optimizing waveguide positions under negligible coupling improves the sum rate by approximately 39\% over fixed uniform deployment, while further allowing closely spaced, coupling-aware deployment provides an additional 19\% improvement over the minimum-spacing benchmark, corresponding to an overall improvement of approximately 65\%.
\end{enumerate}

\section{System Model}\label{sec:II}
\begin{figure}
    \centering
    \includegraphics[width=0.76\linewidth]{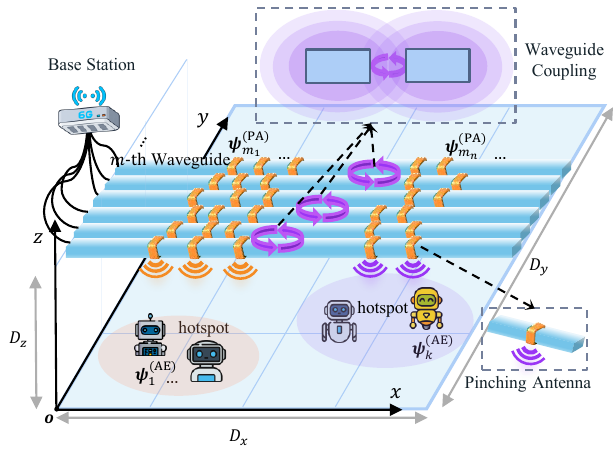}
    \caption{Illustration of the considered CWPASS, where an access point feeds $M$ coupled single-mode dielectric waveguides, each hosting $N$ PAs, to serve physical agents distributed within multiple hotspots.}
    \label{fig:system_model}
\end{figure}

We consider a coupled-waveguide pinching-antenna system, as illustrated in Fig.~\ref{fig:system_model}.
The system comprises $M$ single-mode dielectric waveguides, each connected to an independent radio-frequency chain and hosting $N$ PAs.
The dielectric waveguides and PAs share an identical rectangular cross section with dimensions $a\times b$, where the refractive indices of the waveguide core and cladding are denoted by $n_1$ and $n_0$, respectively.
An internal attenuation coefficient $\alpha_{\mathrm W}$ represents the propagation loss per unit length within the waveguides.

\subsection{Spatial Configuration}


The service region is a 3D volume of size $D_x\times D_y\times D_z$.
Any point in the Cartesian coordinate system is represented by the coordinate tuple $\bm{\psi}=(x,y,z)$.

\subsubsection{Physical Agents (AE)}

The position of the $k$-th physical agent ($k=1,\ldots,K$) is $\bm{\psi}^{(\mathrm{AE})}_k\triangleq\big(x^{(\mathrm{AE})}_k,y^{(\mathrm{AE})}_k,0\big)$.
The $K$ single-antenna physical agents are uniformly distributed within several hotspots in the service region.

\subsubsection{Waveguides (WG)}

The $M$ waveguides are placed horizontally at height $D_z$, oriented parallel to the $x$-axis, with potentially nonuniform spacing along the $y$-axis.
Each waveguide has length $D_x$.
The feed point of the $m$-th waveguide ($m=1,\ldots,M$) is denoted by $\bm{\psi}^{(\mathrm{WG})}_m\triangleq\left(x^{(\mathrm{WG})}_m,y^{(\mathrm{WG})}_m,z^{(\mathrm{WG})}_m\right)$, where $z^{(\mathrm{WG})}_m=D_z$ and $x^{(\mathrm{WG})}_m=0$.

\subsubsection{PAs}

On waveguide $m$, $N$ PAs are deployed.
The $n$-th PA on waveguide $m$ is indexed by $m_n$, and its
radiation center is denoted by
$\bm{\psi}^{(\mathrm{PA})}_{m_n}
\triangleq
\big(
x^{(\mathrm{PA})}_{m_n},
y^{(\mathrm{PA})}_{m_n},
z^{(\mathrm{PA})}_{m_n}
\big)$,
where
$y^{(\mathrm{PA})}_{m_n}=y^{(\mathrm{WG})}_m$
and
$z^{(\mathrm{PA})}_{m_n}=D_z$.
Each PA extracts a fraction of the local guided signal, while
the remaining signal continues to propagate along the coupled
waveguides.

\subsection{Signal Model in CWPASS}

Let $\bm{s}_0\in\mathbb{C}^{K\times1}$ denote the vector of
independent data symbols intended for the $K$ physical agents, with
$\mathbb{E}\{\bm{s}_0\bm{s}_0^{\mathsf H}\}=\bm I_K$.
The digital precoder $\bm W\in\mathbb{C}^{M\times K}$ maps the
$K$ data streams to the $M$ waveguide inputs as
\begin{equation}
\bm s=\bm W\bm s_0,\qquad
\operatorname{tr}(\bm W\bm W^{\mathsf H})\le1,
\label{e:s}
\end{equation}
where the $m$-th entry $s_m$ of $\bm s$ denotes the excitation
launched into the $m$-th waveguide.

After entering the waveguides, the signals undergo
inter-waveguide coupling and distributed PA extraction
before propagating through free space to the physical agents.
Accordingly, the received signal vector
$\bm\xi \in\mathbb{C}^{K\times1}$ is given by
\begin{equation}
\begin{aligned}
\bm \xi
&=
\sqrt{P}\bm H\bm W\bm s_0+\bm\nu
\\
&\triangleq
\sqrt{P}
\bm H^{(\mathrm{P}\to\mathrm{A})}
\bm H^{(\mathrm{W}\to\mathrm{P})}
\big(\bm H^{(\mathrm{CP})}\big)
\bm W\bm s_0
+\bm\nu,
\end{aligned}
\label{eq:end_to_end}
\end{equation}
where $P$ is the total transmit power,
$\bm\nu\sim\mathcal{CN}(\bm0,\sigma^2\bm I_K)$ is the
additive white Gaussian noise (AWGN), and
$\bm H\in\mathbb{C}^{K\times M}$ is the effective
end-to-end channel matrix.
The channel model in \eqref{eq:end_to_end} consists of
three components:
\begin{itemize}[leftmargin=0.4cm]
    \item \textit{Coupled-propagation channel:}
    $\bm H^{(\mathrm{CP})}\in\mathbb{C}^{M\times M}$
    characterizes the signal transfer among the $M$
    waveguides induced by inter-waveguide electromagnetic
    coupling over a given propagation distance.

    \item \textit{Waveguide-to-PA channel:}
    $\bm H^{(\mathrm{W}\to\mathrm{P})}
    \big(\bm H^{(\mathrm{CP})}\big)
    \in\mathbb{C}^{MN\times M}$
    maps the $M$ waveguide inputs to the $MN$ PA outputs,
    accounting for coupled propagation, waveguide attenuation,
    PA locations, and distributed PA extraction.

    \item \textit{PA-to-agent channel:}
    $\bm H^{(\mathrm{P}\to\mathrm{A})}
    \in\mathbb{C}^{K\times MN}$
    describes free-space propagation from the $MN$ PAs
    to the $K$ physical agents and is specified below.
\end{itemize}

We characterize the free-space propagation from the PAs to
the physical agents. Each PA follows a cosine-power radiation
pattern given by \cite{balanis2005antenna}
\begin{equation}
G_{\mathrm{PA}}(\varphi)
=
\begin{cases}
2(\zeta+1)\cos^\zeta\varphi,
& 0\le\varphi\le\pi/2,\\
0,
& \text{otherwise},
\end{cases}
\label{eq:pa_radiation_pattern}
\end{equation}
where $\zeta\ge0$ controls the radiation directivity and
$\varphi$ denotes the angle measured from the downward boresight.
For notational convenience, we index the $MN$ PAs by
$q=1,\ldots,MN$, with $\bm\psi_q^{(\mathrm{PA})}$
denoting the position of the corresponding PA.
For physical agent $k$ and PA $q$, the corresponding radiation
angle is determined by
\begin{equation}
\cos\varphi_{k,q}
=
\frac{
z_q^{(\mathrm{PA})}
}{
\left\|
\bm\psi_k^{(\mathrm{AE})}
-
\bm\psi_q^{(\mathrm{PA})}
\right\|
}.
\label{eq:pa_user_angle}
\end{equation}
To retain the spherical-wave amplitude and phase variations across
the distributed PAs without invoking a plane-wave approximation,
the corresponding PA-to-agent channel coefficient is modeled as
\begin{equation}
h_{k,q}^{(\mathrm{P}\to\mathrm{A})}
\!=\!
\frac{
\lambda
\sqrt{
G_{\mathrm{PA}}(\varphi_{k,q})
}}{
4\pi
\left\|
\bm\psi_k^{(\mathrm{AE})}
\!-\!
\bm\psi_q^{(\mathrm{PA})}
\right\|
}
e^{
-jk_0
\left\|
\bm\psi_k^{(\mathrm{AE})}
-
\bm\psi_q^{(\mathrm{PA})}
\right\|},
\label{eq:pa_user_channel}
\end{equation}
where $k_0=2\pi/\lambda$ is the free-space wavenumber,
$\lambda=c/f$ is the wavelength at operating frequency $f$,
and $c$ is the speed of light in vacuum.
Collecting the channel coefficients of all PAs, the
PA-to-agent channel matrix is constructed as
\begin{equation}
\bm H^{(\mathrm{P}\to\mathrm{A})}
=
\begin{bmatrix}
h_{1,1}^{(\mathrm{P}\to\mathrm{A})}
&
\cdots
&
h_{1,MN}^{(\mathrm{P}\to\mathrm{A})}
\\
\vdots
&
\ddots
&
\vdots
\\
h_{K,1}^{(\mathrm{P}\to\mathrm{A})}
&
\cdots
&
h_{K,MN}^{(\mathrm{P}\to\mathrm{A})}
\end{bmatrix}
\in\mathbb C^{K\times MN}.
\label{eq:pa_user_channel_matrix}
\end{equation}

Let $\bm h_k^{\mathsf H}$ denote the $k$-th row of $\bm H$,
and let $\bm w_k$ denote the $k$-th column of $\bm W$.
The received signal at physical agent $k$ is
\begin{equation}
\xi_k
=
\sqrt{P}\bm h_k^{\mathsf H}\bm w_k s_{0,k}
+
\sqrt{P}\sum_{i\ne k}
\bm h_k^{\mathsf H}\bm w_i s_{0,i}
+\nu_k,
\label{eq:user_received}
\end{equation}
where the first term is the desired signal and the second
term represents multi-agent interference.
Accordingly, treating multi-agent interference as noise,
the achievable rate of physical agent $k$ is
\begin{equation}
R_k
=
\log_2\!\left(
1+
\frac{
P\left|\bm h_k^{\mathsf H}\bm w_k\right|^2
}{
P\sum_{i\ne k}
\left|\bm h_k^{\mathsf H}\bm w_i\right|^2
+\sigma^2
}
\right).
\label{eq:k_rate}
\end{equation}

The above signal model reveals two key characteristics
of CWPASS:
\begin{itemize}[leftmargin=0.4cm]
    \item \textit{Coupling-dependent signal propagation:}
    Unlike conventional PASS with independent waveguides,
    CWPASS allows guided signals to exchange energy among
    different waveguides.
    Since coupled propagation and PA extraction occur
    successively along the propagation direction,
    $\bm H^{(\mathrm{CP})}$ is incorporated into the
    waveguide-to-PA channel rather than appearing as
    an independent cascaded channel.

    \item \textit{Coupling-induced analog beamforming:}
    The interaction between inter-waveguide coupling
    and distributed PA extraction determines the
    amplitudes and phases of the signals radiated by
    different PAs.
    This provides an additional means of shaping
    the effective multi-agent channels beyond digital
    precoding at the waveguide inputs.
\end{itemize}
The coupled-propagation and waveguide-to-PA channels
are derived in the next section.

\section{Evanescent Coupling Characterization}\label{sec:III}

In this section, we characterize the evanescent coupling and guided-wave propagation in CWPASS in three steps.
First, we characterize the guided mode and its evanescent field penetrating neighboring dielectric waveguides.
Second, we derive the induced polarization current and project it onto the guided mode of the receiving waveguide to obtain the pairwise inter-waveguide coupling coefficient.
Finally, we establish the coupled-propagation model and incorporate PA extraction to construct the waveguide-to-PA channel of CWPASS.

\subsection{Evanescent Field Penetration}
\label{subsec:guided_evanescent_fields}

We first characterize the electromagnetic field carried by an
individual dielectric waveguide before introducing the interaction
among different waveguides. This provides the physical basis for the
inter-waveguide coupling coefficients derived subsequently.

To characterize the guided electromagnetic field supported by each
dielectric waveguide, we start from the vector wave equation governing
the electric field,
\begin{equation}
\nabla\times\nabla\times\bm{\mathcal E}
-
\omega^2\psi_0\varepsilon(y,z)\bm{\mathcal E}
=
\bm 0 ,
\label{eq:vector_wave_equation}
\end{equation}
where $\omega=2\pi f$ is the angular frequency with $f$ being
the operating frequency, $\psi_0$ and $\varepsilon_0$ are the
vacuum permeability and permittivity, respectively, and
$\varepsilon(y,z)$ describes the dielectric distribution over the
waveguide cross section.
For the $m$-th isolated waveguide, the
rectangular core is $\mathcal S_m\!=\!\{\!(y,z)\!:\!
|y\!-\!y_m^{(\mathrm{WG})}|\!\le \!\frac{a}{2},
|z\!-\!D_z|\!\le\! \frac{b}{2}\!\}$, with permittivity $\varepsilon_0n_1^2$ inside
the core and $\varepsilon_0n_0^2$ in the surrounding cladding.

The dielectric waveguide is uniform along the $x$-direction, and
therefore the permittivity distribution $\varepsilon(y,z)$ is
independent of $x$. Consequently, the wave equation in
\eqref{eq:vector_wave_equation} admits guided-mode solutions with a
separable longitudinal dependence. For the $m$-th isolated
waveguide, the fundamental guided mode can therefore be written as
\begin{equation}
\bm{\mathcal E}_m^{(\mathrm{iso})}(x,y,z)
=
\bm{\mathcal E}_m^{(\mathrm{co})}(y,z)e^{-j\beta_gx},
\label{eq:isolated_guided_mode}
\end{equation}
where $\bm{\mathcal E}_m^{(\mathrm{co})}$ is the transverse modal field and
$\beta_g$ is the corresponding longitudinal propagation constant.
Substituting \eqref{eq:isolated_guided_mode} into
\eqref{eq:vector_wave_equation} reduces the three-dimensional wave
equation to a Maxwell eigenvalue problem over the waveguide cross
section, from which $\bm{\mathcal E}_m^{(\mathrm{co})}$ and $\beta_g$ are
determined.

For an open rectangular dielectric waveguide, the transverse
vector field generally does not admit a simple separable
closed form. We therefore adopt the Marcatili approximation,
under which the fundamental mode has a dominant electric
field component along the $y$-direction, while the weaker
field components in the corner regions are neglected
\cite{marcatili1969dielectric}. The electric field inside
the core of waveguide $m$ is 
\cite{okamoto2021fundamentals}
\begin{equation}
\hspace{-0.2cm}\bm{\mathcal E}_m^{(\mathrm{co})}(y,z)
\!\approx\!
\mathcal E_0
\cos\!\left[\!
k_y\!\left(y\!-\!y_m^{(\mathrm{WG})}\right)\!
\right]
\cos\!\left[
k_z(z\!-\!D_z)
\right]\bm e_y,
\label{eq:core_mode_field}
\end{equation}
where $\bm e_y=(0,1,0)^{\top}$, $(y,z)\in\mathcal S_m$, and $\mathcal E_0$ is the normalization factor under the scalar power convention defined below. The transverse wavenumbers $k_y$ and $k_z$ are parameterized by the slab-based dispersion relations
\begin{equation}
\label{eq:transverse_wavenumber_y}
   \tan\!\left(\frac{k_ya}{2}\right)
\!=\!\frac{n_1^2\alpha_y}{n_0^2k_y}, \quad 
\tan\!\left(\frac{k_zb}{2}\right)
\!=\!
\frac{
\sqrt{(\frac{\omega}{c})^2(n_1^2-n_0^2)-k_z^2}
}{
k_z
},
\end{equation}
where
$\alpha_y\triangleq
\sqrt{(\frac{\omega}{c})^2(n_1^2-n_0^2)-k_y^2}$
is the lateral evanescent decay constant. The dimensionless
quantities $k_ya/2$ and $k_zb/2$ represent the transverse phase
accumulated from the center of the waveguide to the corresponding
core--cladding interface. The longitudinal propagation constant is
$\beta_g=
\sqrt{(\omega n_1/c)^2-k_y^2-k_z^2}$.

\begin{prop}[Unit-Power Modal Normalization]
\label{prop:modal_normalization}
Under the Marcatili approximation, the normalization factor
$\mathcal E_0$ that makes the isolated fundamental mode carry unit
guided power is given by
\begin{align}\label{eq:modal_normalization_factor}
\mathcal E_0=
&\Bigg\{
\frac{2\omega\psi_0}{\beta_g}
\left[
\frac{a}{2}
+\frac{\sin(k_ya)}{2k_y}
+\frac{\cos^2(\frac{k_ya}{2})}{\alpha_y}
\right]^{-1}
\\
\times&
\left[
\frac{b}{2}
+\frac{\sin(k_zb)}{2k_z}
+
\frac{
\cos^2(\frac{k_zb}{2})
}{
\sqrt{
(\frac{\omega}{c})^2(n_1^2-n_0^2)-k_z^2
}
}
\right]^{-1}
\Bigg\}^{1/2}.\nonumber
\end{align}
\end{prop}

\begin{proof}
Please see the supplementary material.
\end{proof}

Using the unit-power normalized fundamental mode above, the
guided electric field associated with the excitation
$s_m$ in the $m$-th waveguide is
\begin{equation}
\bm{\mathcal E}_m^{(\mathrm{WG})}(x,y,z)
=
\bm{\mathcal E}_m^{(\mathrm{co})}(y,z)s_m
e^{-\frac{\alpha_{\mathrm W}}{2}x}
e^{-j\beta_gx}.
\label{eq:guided_field_CWPASS}
\end{equation}
Here,
$e^{-\frac{\alpha_{\mathrm W}}{2}x}$ accounts for the internal waveguide
attenuation and $e^{-j\beta_gx}$ represents the guided-wave
phase accumulation.
We next characterize the evanescent field extending from one
waveguide to another.

\begin{thm}[Pairwise Evanescent Field]
\label{thm:pairwise_evanescent_field}
Consider two distinct waveguides $m$ and $m'$, and without loss of
generality assume
$y_{m'}^{(\mathrm{WG})}>y_m^{(\mathrm{WG})}$.
Under the adopted guided-mode model, the incident tail of waveguide $m$, evaluated in the region occupied by core $m'$, is given by
\begin{align}
\label{eq:evanescent_field_mm}
&\bm{\mathcal E}_{m\rightarrow m'}^{(\mathrm{ev})}(x,y,z)
=
\mathcal E_0 s_m
e^{-\frac{\alpha_{\mathrm W}}{2}x}
e^{-j\beta_gx}
\cos\!\left(\frac{k_ya}{2}\right)
\\
&\times
\exp\!\left[
-\alpha_y
\left(
\frac{a}{2}
+y-y_{m'}^{(\mathrm{WG})}
\right)
\right]
\cos\!\left[
k_z(z-D_z)
\right]
\nonumber \\
&\times
\exp\!\left[
-\alpha_y
\left(
\left|
y_m^{(\mathrm{WG})}
-
y_{m'}^{(\mathrm{WG})}
\right|-a
\right)
\right] \bm e_y, \quad(y,z)\in\mathcal S_{m'}.\nonumber
\end{align} 
\end{thm}

\begin{proof}
Please see the supplementary material.
\end{proof}

\begin{figure}
    \centering
    \includegraphics[width=0.8\linewidth]{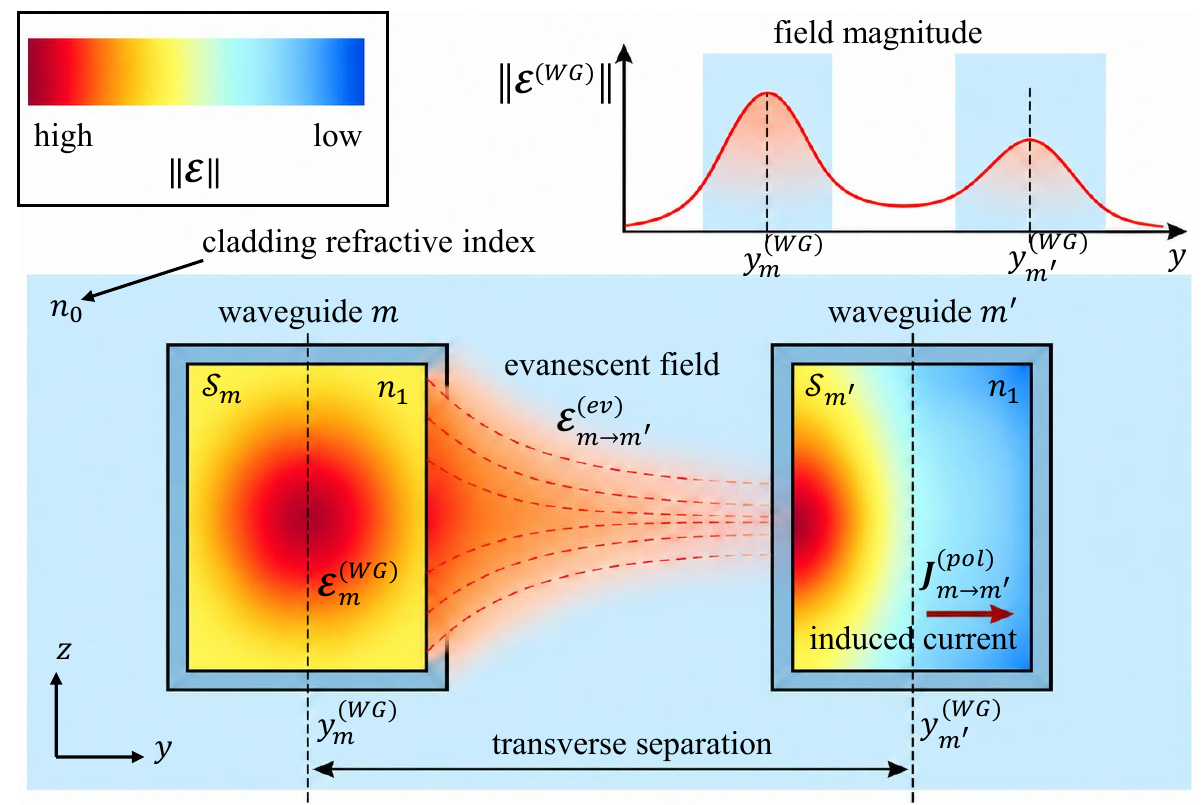}
    \caption{Evanescent-field interaction between two neighboring dielectric waveguides. }
     \label{fig:evanescent_field}
\end{figure}

As illustrated in Fig.~\ref{fig:evanescent_field},
Theorem~\ref{thm:pairwise_evanescent_field}
shows that the guided field extends into the
surrounding cladding and decays exponentially
along the transverse direction.
Its penetration into the neighboring waveguide
core induces polarization currents, providing
the physical basis for inter-waveguide coupling.
Consequently, the interaction weakens as the
waveguide separation increases.

\subsection{Coupling Coefficient}
\label{subsec:modal_projection_coupling}

The evanescent field derived in
Theorem~\ref{thm:pairwise_evanescent_field} penetrates into the
dielectric core of waveguide $m'$ and induces a polarization
response due to the permittivity contrast between the core and the
surrounding cladding. We first characterize the resulting equivalent
polarization current and then project it onto the fundamental guided
mode of waveguide $m'$ to determine the pairwise coupling
coefficient.

\begin{cor}[Induced Polarization Current]
\label{lem:polarization_current}
The evanescent field of waveguide $m$ induces
the following polarization current density
within the core $\mathcal S_{m'}$:
\begin{equation}
\bm J_{m\rightarrow m'}^{(\mathrm{pol})}(x,y,z)
=
j\omega\varepsilon_0
\left(n_1^2\!-\!n_0^2\right)
\bm{\mathcal E}_{m\rightarrow m'}^{(\mathrm{ev})}(x,y,z),
\label{eq:polarization_current_mm}
\end{equation}
\end{cor}


The polarization current in \eqref{eq:polarization_current_mm}
acts as a distributed source that excites the fundamental mode
of waveguide $m'$ through modal overlap. For the unit-power
normalized mode in \eqref{eq:unit_power_condition}, the forward
excitation amplitude is given by \cite{pozar2021microwave}
\begin{equation}
A_{m\rightarrow m'}
\!=\!
-\frac{1}{4}\!
\iiint_V
\!\left[\!
\bm{\mathcal E}_{m'}^{(\mathrm{iso})}(x,y,z)
\!\right]^*\!\!\!\cdot\!
\bm J_{m\rightarrow m'}^{(\mathrm{pol})}(x,y,z)
dV,\!\!\!
\label{eq:standard_modal_excitation}
\end{equation}
where $V$ denotes the integration volume within the core of
waveguide $m'$, and $A_{m\rightarrow m'}$ denotes the excited
forward fundamental-mode amplitude in waveguide $m'$.
Since the polarization current is distributed along the
$x$-direction, the excitation per unit length becomes
\begin{align}\label{eq:modal_current_projection}
&\frac{dA_{m\rightarrow m'}}{dx}
=
-\frac{1}{4}e^{j\beta_gx}
\iint_{\mathcal S_{m'}}
\left[
\bm{\mathcal E}_{m'}^{(\mathrm{co})}
\right]^*
\cdot
\bm J_{m\rightarrow m'}^{(\mathrm{pol})}
\,dy\,dz
\\
&\overset{(a)}{=}
-j\frac{\omega\varepsilon_0(n_1^2-n_0^2)}{4}
e^{j\beta_gx}
\iint_{\mathcal S_{m'}}
\left[
\bm{\mathcal E}_{m'}^{(\mathrm{co})}
\right]^*
\cdot
\bm{\mathcal E}_{m\rightarrow m'}^{(\mathrm{ev})}
\,dy\,dz, \nonumber
\end{align}
where step (a) follows from
\eqref{eq:polarization_current_mm} under the
dominant-field approximation.
Since $\bm{\mathcal E}_{m\rightarrow m'}^{(\mathrm{ev})}$
contains the longitudinal factor
$s_m e^{-\alpha_{\mathrm W}x/2}e^{-j\beta_gx}$,
the guided-wave phases cancel. We thus define the
coupling coefficient $\mu_{mm'}$ through
\begin{equation}
\frac{dA_{m\rightarrow m'}}{dx}
\triangleq
-j\mu_{mm'}s_m e^{-\frac{\alpha_{\mathrm W}}{2}x},
\label{eq:coupling_coefficient_definition}
\end{equation}
where $\mu_{mm'}$ characterizes the mode conversion from waveguide
$m$ to waveguide $m'$ per unit propagation length.

\begin{thm}[Pairwise Inter-Waveguide Coupling]
\label{thm:pairwise_coupling}
For two nonoverlapping identical rectangular waveguides, evaluating \eqref{eq:modal_current_projection} with the scalar envelope gives the coefficient model
\begin{align}\label{eq:pairwise_coupling_closed_form}
&\mu_{mm'} \!\left(d_{mm'}^{(\mathrm{WG})}\right)
\!\approx\!
\frac{
|\mathcal E_0|^2\alpha_y
\cos^2\!\left(\!\frac{k_ya}{2}\!\right)
}{
4\omega\psi_0
}
\left(
\frac{b}{2}
\!+\!
\frac{\sin(k_zb)}{2k_z}
\right)
\\
&\times
\left[
1+\frac{n_1^2}{n_0^2}
+
\left(
\frac{n_1^2}{n_0^2}-1
\right)e^{-\alpha_ya}
\right]
\exp\!\left[
-\alpha_y
\left(
d_{mm'}^{(\mathrm{WG})}-a
\right)
\right],\nonumber
\end{align}
where the transverse distance between waveguides $m$ and $m'$ is
defined as
$d_{mm'}^{(\mathrm{WG})}
\triangleq
\left|
y_m^{(\mathrm{WG})}
-
y_{m'}^{(\mathrm{WG})}
\right|$.
\end{thm}

\begin{proof}
Without loss of generality, assume
$y_{m'}^{(\mathrm{WG})}>y_m^{(\mathrm{WG})}$.
Substituting \eqref{eq:polarization_current_mm},
\eqref{eq:core_mode_field}, and
\eqref{eq:evanescent_field_mm} into
\eqref{eq:modal_current_projection}
and separating the transverse modal-overlap
integrals yields
\begin{align}\label{eq:coupling_separated_integrals}
\mu_{mm'}
&\!=\!
\frac{
\omega\varepsilon_0(n_1^2\!-\!n_0^2)
|\mathcal E_0|^2
}{4}
\cos\!\left(\!\frac{k_ya}{2}\!\right) \!
e^{-\alpha_y
\left(
\left|
y_m^{(\mathrm{WG})}
\!-y_{m'}^{(\mathrm{WG})}
\right|
-\frac{a}{2}
\right)}
\nonumber \\
&\times
\int_{y_{m'}^{(\mathrm{WG})}-\frac{a}{2}}
     ^{y_{m'}^{(\mathrm{WG})}+\frac{a}{2}}
e^{-\alpha_y(y-y_{m'}^{(\mathrm{WG})})}
\cos\!\left[
k_y(y-y_{m'}^{(\mathrm{WG})})
\right]dy
\nonumber \\
&\times
\int_{D_z-\frac{b}{2}}^{D_z+\frac{b}{2}}
\cos^2\!\left[k_z(z-D_z)\right]dz .
\end{align}
The vertical integral evaluates to
$b/2+\sin(k_zb)/(2k_z)$, while the lateral
integral is evaluated using the boundary
condition in \eqref{eq:transverse_wavenumber_y}.
Applying
$k_y^2+\alpha_y^2
=(\omega/c)^2(n_1^2-n_0^2)$
and $\psi_0\varepsilon_0=c^{-2}$,
and collecting the exponential terms,
yields \eqref{eq:pairwise_coupling_closed_form}.
\end{proof}

Theorem~\ref{thm:pairwise_coupling} shows that the coupling
coefficient consists of a common modal factor determined by the
waveguide structure and an exponential factor determined by the
transverse separation between the two waveguides. Therefore, for
fixed waveguide parameters, the inter-waveguide coupling strength
can be directly controlled through the waveguide spacing and decays
exponentially as the separation increases.

\subsection{Coupled Propagation}
\label{subsec:coupled_channel}

Collecting the pairwise coupling coefficients derived in
Theorem~\ref{thm:pairwise_coupling}, the coupling matrix of the
$M$ waveguides is written as
\begin{equation}
\bm C=
\begin{bmatrix}
0 & \mu_{12} & \mu_{13} & \cdots & \mu_{1M}\\
\mu_{12} & 0 & \mu_{23} & \cdots & \mu_{2M}\\
\mu_{13} & \mu_{23} & 0 & \cdots & \mu_{3M}\\
\vdots & \vdots & \vdots & \ddots & \vdots\\
\mu_{1M} & \mu_{2M} & \mu_{3M} & \cdots & 0
\end{bmatrix},
\label{eq:coupling_matrix}
\end{equation}
where $\mu_{mm'}=\mu_{m'm}$ follows from reciprocity between the
identical waveguides.

\begin{figure}[!tb]
\centering
    \begin{subfigure}[!tb]{0.415\linewidth}
        \centering
        \includegraphics[width=0.72\linewidth]{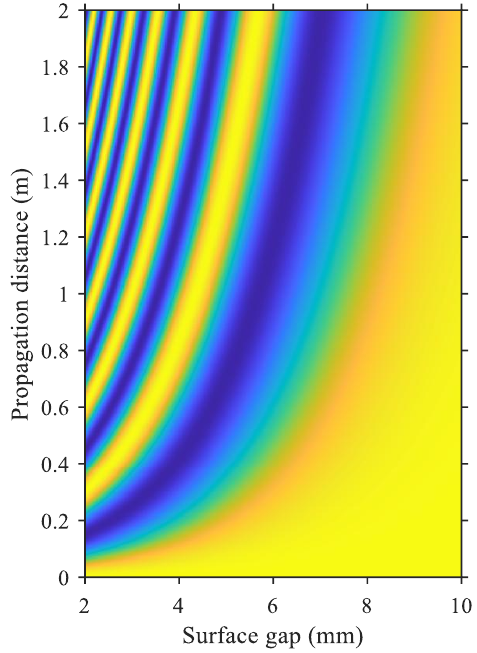}
        \caption{}\label{f:map1}
    \end{subfigure}
    \begin{subfigure}[!tb]{0.505\linewidth}
        \centering
        \includegraphics[width=0.72\linewidth]{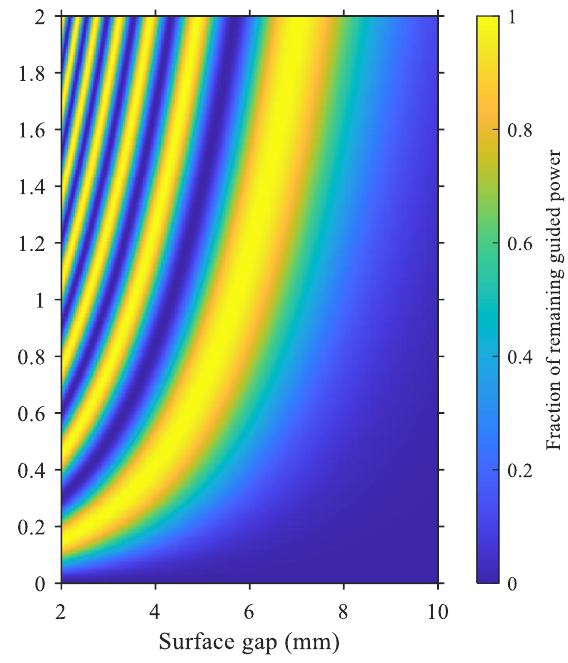}
        \caption{}\label{f:map2}
    \end{subfigure}

\caption{Power exchange in two identical guides with only the first guide excited and no PA extraction: (a) the excited guide; (b) the adjacent guide.}
\label{fig:power_map}
\end{figure}

Since the waveguide structure is invariant along the
$x$-direction, $\bm C$ is independent of the propagation distance.
For a longitudinal propagation distance $\Delta x$, the
coupled-propagation channel is
\begin{equation}
\bm H^{(\mathrm{CP})}(\Delta x)
=
e^{-(\frac{\alpha_{\mathrm W}}{2}+j\beta_g)\Delta x}
\exp\!\left(-j\bm C\Delta x\right),
\label{eq:coupled_propagation_channel}
\end{equation}
which captures guided-wave attenuation, phase
accumulation, and inter-waveguide energy exchange.
Fig.~\ref{fig:power_map} illustrates the periodic
power exchange between two coupled waveguides.
Increasing the inter-waveguide gap exponentially
weakens coupling and lengthens the power-transfer
period, implying that stronger coupling does not
necessarily yield greater power transfer at a
given propagation distance.

We next incorporate sequential PA extraction.
For PA $m_n$, the coupling length $\tau_{m_n}$
determines the extracted and remaining guided-field
amplitude factors $\sin(\kappa\tau_{m_n})$ and
$\cos(\kappa\tau_{m_n})$, respectively
\cite{wang2025modeling}, where $\kappa$ denotes
the waveguide-to-PA coupling coefficient.
All $MN$ PAs are indexed by $q=1,\ldots,MN$
in ascending longitudinal order,
$x_1^{(\mathrm{PA})}\le\cdots\le
x_{MN}^{(\mathrm{PA})}$.
Each index retains its hosting waveguide and
coupling length $\tau_q$, with the same ordering
applied to the columns of
$\bm H^{(\mathrm{P}\to\mathrm{A})}$.

\begin{thm}[Waveguide-to-PA Channel]
\label{thm:waveguide_to_pa_channel}
With all $MN$ PAs ordered according to their longitudinal
coordinates, let $x_0^{(\mathrm{PA})}=0$.
If the $q$-th PA is located on waveguide $m$, the
corresponding row of the waveguide-to-PA channel is
\begin{equation}
\begin{aligned}
&
\left[
\bm H^{(\mathrm{W}\to\mathrm{P})}
\right]_{q,:}
=
\sin(\kappa\tau_q)
\Bigg[
\bm H^{(\mathrm{CP})}
\left(
x_q^{(\mathrm{PA})}
-
x_{q-1}^{(\mathrm{PA})}
\right)
\\
&\qquad\times
\prod_{i=q-1}^{1}
\left[
\bm D_i
\bm H^{(\mathrm{CP})}
\left(
x_i^{(\mathrm{PA})}
-
x_{i-1}^{(\mathrm{PA})}
\right)
\right]
\Bigg]_{m,:},
\end{aligned}
\label{eq:waveguide_to_pa_channel}
\end{equation}
where
$\bm D_i\in\mathbb C^{M\times M}$ is a diagonal matrix
whose diagonal entries are all one except for the entry
corresponding to the waveguide hosting PA $i$, which is
$\cos(\kappa\tau_i)$, i.e.,
\[
\bm D_i
\triangleq
\operatorname{diag}
\left(
1,\ldots,
\cos(\kappa\tau_i),
\ldots,1
\right).
\]
The product in \eqref{eq:waveguide_to_pa_channel} is taken in
descending order from $i=q-1$ to $i=1$.
\end{thm}

Theorem~\ref{thm:waveguide_to_pa_channel} shows that each row of
$\bm H^{(\mathrm{W}\to\mathrm{P})}$
is formed by the ordered cascade of coupled propagation and the
residual-field factors introduced by all preceding PAs.
The current PA then extracts the field amplitude
$\sin(\kappa\tau_q)$ from its hosting waveguide.
Together with the PA-to-agent channel in
\eqref{eq:pa_user_channel_matrix}, the waveguide-to-PA
channel in \eqref{eq:waveguide_to_pa_channel} completes
the effective end-to-end channel $\bm H$ in
\eqref{eq:end_to_end}.

\section{Coupling-Aware Rate Upper Bounds}\label{sec:IV}
\begin{figure*}[!t]
\normalsize

\setcounter{CWtempeq}{\value{equation}}

\setcounter{equation}{\getrefnumber{eq:effective_channel_section}}

\begin{equation}
\begin{aligned}
\|\bm h_k\|^2
\approx&
\underbrace{
\frac{\lambda^2(\zeta+1)D_z^\zeta}{8\pi^2N}
\sum_{m=1}^{M}\sum_{n=1}^{N}
\frac{
e^{-\alpha_{\mathrm W}x_{m_n}^{(\mathrm{PA})}}
}{
r_{k,m_n}^{\zeta+2}
}
}_{\text{power accumulation}}
+\!
\underbrace{
\begin{aligned}
&\frac{\lambda^2(\zeta+1)D_z^\zeta}{4\pi^2N}
\operatorname{Re}\!\Bigg\{\!
\sum_{1\le n<n'\le N}
\sum_{m=1}^{M}\sum_{m'=1}^{M}
\frac{
e^{-\frac{\alpha_{\mathrm W}}{2}
\left(
x_{m_n}^{(\mathrm{PA})}
+
x_{m'_{n'}}^{(\mathrm{PA})}
\right)}
}{
\left(
r_{k,m_n}r_{k,m'_{n'}}
\right)^{\frac{\zeta+2}{2}}
}
\\[-1mm]
&\hspace{7mm}\times
e^{-j\left[
k_0
\left(
r_{k,m_n}-r_{k,m'_{n'}}
\right)
+
\beta_g
\left(
x_{m_n}^{(\mathrm{PA})}
-
x_{m'_{n'}}^{(\mathrm{PA})}
\right)
\right]}
\left[
e^{-j\bm C(L_n-L_{n'})}
\right]_{m,m'}
\!\Bigg\}
\end{aligned}
}_{\text{coherent combining}} .
\end{aligned}
\label{eq:self_cross_coupling_gain}
\end{equation}

\setcounter{equation}{\value{CWtempeq}}

\vspace{-1mm}
\end{figure*}

In this section, we characterize the coupling-aware rate bounds of
CWPASS in two steps. First, we investigate coherent combining across
PA extraction sections and establish its dependence on inter-waveguide
coupling. Second, we derive an achievable-rate upper bound and analyze
its scaling with the number of waveguides.

\label{subsec:performance_bound}

\subsection{Inter-Section Coherence}
\label{subsec:inter_section_coherence}

We first characterize the effective channel gain under distributed
PA extraction and identify the inter-section coherent contributions.
We adopt an offline equal-quota extraction design, under which
the PA coupling lengths are predetermined as
$\tau_{m_n}
=\kappa^{-1}\arcsin\sqrt{(N+1-n)^{-1}}$
and require no real-time adjustment
\cite{zhang2026directional}.

To characterize the coherent interaction among distributed PAs,
we adopt a local scale-separation model.
For the $n$-th PA group, let
$\underline{x}_n=\min_m x_{m_n}^{(\mathrm{PA})}$
and
$\overline{x}_n=\max_m x_{m_n}^{(\mathrm{PA})}$.
The groups occupy nonoverlapping short sections satisfying
$\overline{x}_n<\underline{x}_{n+1}$ and
$\|\bm C\|_2(\overline{x}_n-\underline{x}_n)\ll1$.
Accordingly, the additional coupling evolution within each
short section is neglected, while the accumulated coupling
propagation outside these sections is retained.

Substituting the prescribed coupling lengths into
\eqref{eq:waveguide_to_pa_channel} under this approximation yields
\begin{equation}
\begin{aligned}
\bm H^{(\mathrm{W}\to\mathrm{P})}
\!\approx \!\frac{1}{\sqrt N}\bm P_x\!\!
\begin{bmatrix}
\!\operatorname{diag}_{m=1}^{M}
\left(
e^{-\left(\frac{\alpha_{\mathrm W}}{2}+j\beta_g\right)
x_{m_1}^{(\mathrm{PA})}}
\right)e^{-j\bm C L_1}\!\\
\vdots\\
\!\operatorname{diag}_{m=1}^{M}
\left(
e^{-\left(\frac{\alpha_{\mathrm W}}{2}+j\beta_g\right)
x_{m_N}^{(\mathrm{PA})}}
\right)e^{-j\bm C L_N}\!
\end{bmatrix},
\end{aligned}
\label{eq:waveguide_pa_section_approx}
\end{equation}
where $\bm P_x\in\{0,1\}^{MN\times MN}$ is a permutation
matrix that rearranges the rows from the section-wise
waveguide ordering to the longitudinal PA ordering, and
$L_n=\underline{x}_n-
\sum_{\ell=1}^{n-1}
(\overline{x}_\ell-\underline{x}_\ell)$
denotes the effective coupled-propagation length to the
$n$-th extraction section, excluding the lengths of
preceding extraction sections.

Combining \eqref{eq:waveguide_pa_section_approx} with
\eqref{eq:pa_user_channel}, the effective channel of
physical agent $k$ is given by
\begin{align}
\label{eq:effective_channel_section}
&\bm h_k^{\mathsf H}
=
\frac{\lambda\sqrt{2(\zeta+1)}D_z^{\frac{\zeta}{2}}}
{4\pi\sqrt N}
\sum_{n=1}^{N}\sum_{m=1}^{M}
\frac{
e^{-\left(\frac{\alpha_{\mathrm W}}{2}
+j\beta_g\right)x_{m_n}^{(\mathrm{PA})}}
}{
r_{k,m_n}^{\frac{\zeta+2}{2}}
}
\nonumber\\
&\times
e^{-jk_0r_{k,m_n}}
\left[e^{-j\bm C L_n}\right]_{m,:}
\triangleq \sum_{n=1}^{N} \bm h_{k,n}^{\mathsf H} e^{-j\bm C L_n}, 
\end{align}\addtocounter{equation}{1}
where
$r_{k,m_n}
=\|\bm\psi_k^{(\mathrm{AE})}
-\bm\psi_{m_n}^{(\mathrm{PA})}\|$
denotes the distance from PA $m_n$ to physical agent $k$.
Taking the squared norm of
\eqref{eq:effective_channel_section}
and exploiting the unitarity of $e^{-j\bm C L_n}$ yields
\eqref{eq:self_cross_coupling_gain}.
The decomposition reveals three key properties:
\begin{itemize}
\item \emph{Power accumulation:}
Power accumulation depends on propagation attenuation and
PA-to-agent distances, but not on inter-waveguide coupling.

\item \emph{Coherent combining:}
Coupling controls coherent combining across extraction sections
through $e^{-j\bm C(L_n-L_{n'})}$.
For $N=1$, coupling has no effect on the effective channel gain,
whereas multiple sections allow constructive or destructive combining.

\item \emph{Waveguide spacing design:}
Waveguide spacing should be designed to improve inter-section
coherence rather than simply maximize coupling strength.
\end{itemize}

The coherent cross term in
\eqref{eq:self_cross_coupling_gain} also provides a direct
guideline for selecting the waveguide spacing. For the spacing analysis, we neglect the variation of the
non-coupling propagation response within each short extraction
section.

\begin{cor}[Closed-Form Waveguide Spacing]
\label{cor:optimal_waveguide_spacing}
For $M=2$, suppose the coherent contribution in
\eqref{eq:self_cross_coupling_gain} is dominated by
one extraction-section pair $(n,n')$, $n<n'$,
under the symmetric-supermode approximation.
With exponential coupling
$\mu(d)=\mu(a)e^{-\alpha_y(d-a)}$
and propagation distances evaluated at a fixed
reference geometry, the phase-matched
waveguide-spacing candidates are given by
\begin{align}\label{eq:closed_form_waveguide_spacing}
d_p^\star
=&\frac{1}{\alpha_y}\ln\Bigg\{
\mu(a)(L_{n'}-L_n)
\Bigg[
\frac{1}{2}\sum_{m=1}^{2}
\Big(
k_0(r_{k,m_n}-r_{k,m_{n'}})\nonumber
\\[-1mm]
&\hspace{13mm}
+\beta_g
(x_{m_n}^{(\mathrm{PA})}
-x_{m_{n'}}^{(\mathrm{PA})})
\Big)
+2\pi p
\Bigg]^{-1}
\Bigg\}+a,
\end{align}
where $p\in\mathbb Z$ is chosen such that
the logarithm is well defined and
$d_p^\star$ lies within the feasible
waveguide-spacing interval.
\end{cor}
\begin{proof}
Please see the supplementary material.
\end{proof}

Corollary~\ref{cor:optimal_waveguide_spacing} shows that
waveguide spacing can be selected to align the coupling-induced
phase with the propagation-phase difference between extraction
sections, thereby enhancing coherent combining.
To further bound the inter-section coherence, we consider
$M$ equally spaced waveguides with adjacent spacing $d$,
for which \eqref{eq:pairwise_coupling_closed_form} gives
$C_{mm'}=\mu(d)e^{-\alpha_y(|m-m'|-1)d}$ for $m\ne m'$,
where $\mu(d)$ denotes the adjacent-waveguide coupling coefficient.


\begin{lem}[Bounds on the Coupling-Matrix Norm]
\label{lem:coupling_saturation}
For the equally spaced deployment, the spectral norm of
$\bm C$ satisfies
\begin{align}
\label{eq:coupling_norm_bound}
&\frac{2\mu(d)}{1-e^{-\alpha_y d}}
\left[
1-
\frac{1-e^{-\alpha_y Md}}
{M\left(1-e^{-\alpha_y d}\right)}
\right]
\le
\|\bm C\|_2
\\
&\qquad\le
\frac{\mu(d)}{1-e^{-\alpha_y d}}
\left[
2
-e^{-\alpha_y d
\left\lfloor\frac{M-1}{2}\right\rfloor}
-e^{-\alpha_y d
\left\lceil\frac{M-1}{2}\right\rceil}
\right].\nonumber
\end{align}
\end{lem}

\begin{proof}
Please see the supplementary material.
\end{proof}

Fig.~\ref{fig:coupling_norm} illustrates the impact of
waveguide number and spacing on the coupling strength.
The spectral norm remains within the analytical bounds
in \eqref{eq:coupling_norm_bound}.
For a fixed spacing $d$, $\|\bm C\|_2$ increases with
$M$ and gradually saturates toward
\begin{equation*}
\lim_{M\rightarrow\infty}\|\bm C\|_2
=
\frac{2\mu(d)}{1-e^{-\alpha_y d}}.
\end{equation*}
In contrast, increasing $d$ significantly weakens
inter-waveguide coupling due to the exponential decay
of the evanescent field.

We next examine how inter-waveguide coupling affects
coherent combining across extraction sections.
From \eqref{eq:self_cross_coupling_gain}, the effective
channel gain can be expressed as
\begin{align}\label{eq:h_k_norm}
\|\bm h_k\|^2
=&
\sum_{n=1}^{N}\|\bm h_{k,n}\|^2+
2\sum_{1\le n<n'\le N}
\|\bm h_{k,n}\|\|\bm h_{k,n'}\|
\nonumber\\
&\times
\operatorname{Re}\!\left\{
\frac{
\bm h_{k,n}^{\mathsf H}
e^{j\bm C(L_{n'}-L_n)}
\bm h_{k,n'}
}{
\|\bm h_{k,n}\|\|\bm h_{k,n'}\|
}
\right\}.
\end{align}
The real normalized inner product characterizes the
coherent interaction between two extraction sections.
Its value ranges from $-1$ to $1$, corresponding to
destructive and constructive combining, respectively.
The following proposition establishes an upper bound
on this correlation in terms of the propagation-phase
differences and inter-waveguide coupling.

\begin{prop}[Inter-Section Coherence Bound]
\label{prop:inter_section_coherence}
For $N\ge2$ and $1\le n<n'\le N$, the real normalized
correlation between extraction sections $n$ and $n'$
satisfies
\begin{equation}
\label{eq:inter_section_correlation_bound}
\operatorname{Re}\!\left\{
\frac{
\bm h_{k,n}^{\mathsf H}
e^{j\bm C(L_{n'}-L_n)}
\bm h_{k,n'}
}{
\|\bm h_{k,n}\|\|\bm h_{k,n'}\|
}
\right\}
\le\cos\theta_{k,nn'},
\end{equation}
where
\begin{align}
\label{eq:inter_section_coherence}
\theta_{k,nn'}
\triangleq&
\Bigg[
\min\!\Bigg\{
\frac{\pi}{2},\,
\min_{1\le m\le M}
\arccos\!\Bigg(\cos\!\Big[
k_0(r_{k,m_{n'}}-r_{k,m_n})
\nonumber\\[-1mm]
&+\beta_g\big(
x_{m_{n'}}^{(\mathrm{PA})}
-x_{m_n}^{(\mathrm{PA})}
\big)
\Big]\Bigg)\Bigg\}
-
\frac{\mu(d)(L_{n'}-L_n)}
{1-e^{-\alpha_y d}} \nonumber \\
&\Big[
2-e^{-\alpha_y d
\left\lfloor\frac{M-1}{2}\right\rfloor}
-e^{-\alpha_y d
\left\lceil\frac{M-1}{2}\right\rceil}
\Big]
\Bigg]_+ .
\end{align}
Here, $[t]_+=\max\{t,0\}$, and
$0\le\cos\theta_{k,nn'}\le1$.
\end{prop}

\begin{proof}
By the Cauchy-Schwarz inequality, the
normalized correlation before coupling satisfies
\begin{align}
&\frac{
\operatorname{Re}\{
\bm h_{k,n}^{\mathsf H}\bm h_{k,n'}
\}
}{
\|\bm h_{k,n}\|\|\bm h_{k,n'}\|
}
=
\sum_{m=1}^{M}
\frac{
|[\bm h_{k,n}]_m|
|[\bm h_{k,n'}]_m|
}{
\|\bm h_{k,n}\|\|\bm h_{k,n'}\|
}
\\
&\quad\times
\cos\!\Big[
k_0(r_{k,m_{n'}}-r_{k,m_n})
+\beta_g(
x_{m_{n'}}^{(\mathrm{PA})}
-x_{m_n}^{(\mathrm{PA})})
\Big]
\nonumber\\
&\le\!
\Bigg[\!
\max_{1\le m\le M}
\!\cos\!\Big[
k_0(r_{k,m_{n'}}\!-\!r_{k,m_n})
\!+\!\beta_g(
x_{m_{n'}}^{(\mathrm{PA})}
\!-\!x_{m_n}^{(\mathrm{PA})})
\Big]
\!\Bigg]_+,\nonumber
\end{align}
For nonzero complex vectors, define their
angular separation as
$\langle\bm a,\bm b\rangle
\triangleq
\arccos\big(\frac{\operatorname{Re}\{\bm a^{\mathsf H}\bm b\}}{\|\bm a\|\|\bm b\|}
\big)$.
The coupling-induced angular displacement
is bounded by
\begin{align}
\left\langle
\bm h_{k,n'},
e^{j\bm C(L_{n'}-L_n)}\bm h_{k,n'}
\right\rangle
&\le
\int_0^{L_{n'}-L_n}
\frac{
\left\|
\bm C e^{j\bm Ct}\bm h_{k,n'}
\right\|
}{
\|\bm h_{k,n'}\|
}\,dt
\nonumber\\
&\le
\|\bm C\|_2(L_{n'}-L_n),
\end{align}
Applying the angular triangle inequality,
the initial correlation bound, and
Lemma~\ref{lem:coupling_saturation} gives
\begin{align}
&\left\langle
\bm h_{k,n},
e^{j\bm C(L_{n'}-L_n)}\bm h_{k,n'}
\right\rangle
\\
&\ge
\Bigg[
\left\langle
\bm h_{k,n},\bm h_{k,n'}
\right\rangle
-
\left\langle
\bm h_{k,n'},
e^{j\bm C(L_{n'}-L_n)}\bm h_{k,n'}
\right\rangle
\Bigg]_+
\nonumber\\
&\ge
\Big[
\left\langle
\bm h_{k,n},\bm h_{k,n'}
\right\rangle
-\|\bm C\|_2(L_{n'}-L_n)
\Big]_+
\ge\theta_{k,nn'},\nonumber
\end{align}
Taking the cosine yields
\eqref{eq:inter_section_correlation_bound}.
\end{proof}

\begin{figure}[!t]
    \centering
    \includegraphics[width=0.76\linewidth]{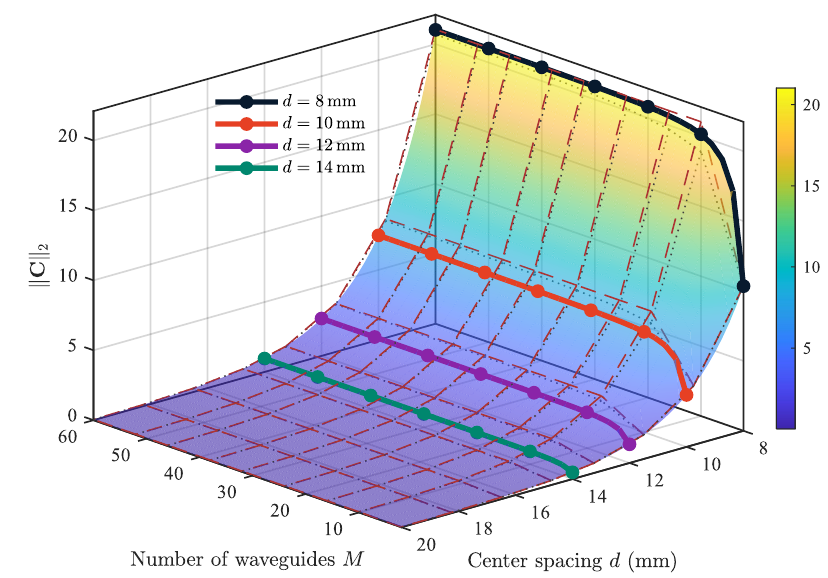}
    \caption{Spectral norm of the inter-waveguide coupling
    matrix versus the number of waveguides $M$ and
    center spacing $d$, with selected spacings highlighted.}
    \label{fig:coupling_norm}
\end{figure}

\subsection{Rate Upper Bound}
\label{subsec:rate_upper_bound}

Combining the coherence bound in
Proposition~\ref{prop:inter_section_coherence}
with the propagation attenuation and deployment
geometry yields the following channel-gain bound.

\begin{thm}[Coupling-Aware Channel-Gain Upper Bound]
\label{thm:coupling_aware_channel_bound}
Under the local scale-separation and equal-quota
extraction model in
\eqref{eq:waveguide_pa_section_approx},
the effective channel gain satisfies
\begin{align}
&\|\bm h_k\|^2
\le
\frac{
\lambda^2(\zeta+1)e^{-\alpha_{\mathrm W}\underline{x}_1}
}{
8\pi^2D_z^2
}
\left[
1+\frac{2}{N}
\sum_{1\le n<n'\le N}
\cos\theta_{k,nn'}
\right]
\nonumber\\
&\times
\sum_{m=1}^{M}
\left[
1+
\frac{
\left(
y_m^{(\mathrm{WG})}
-y_k^{(\mathrm{AE})}
\right)^2
}{D_z^2}
\right]^{-\frac{\zeta+2}{2}}.
\label{eq:effective_channel_geometry_bound}
\end{align}
For $N=1$, \eqref{eq:effective_channel_geometry_bound}
reduces to the single-section bound.
\end{thm}

\begin{proof}
From \eqref{eq:waveguide_pa_section_approx},
the channel gain of each extraction section
satisfies
\begin{align}
&\|\bm h_{k,n}\|^2
=
\frac{\lambda^2(\zeta+1)D_z^\zeta}{8\pi^2N}
\sum_{m=1}^{M}
\frac{
e^{-\alpha_{\mathrm W}x_{m_n}^{(\mathrm{PA})}}
}{
r_{k,m_n}^{\zeta+2}
}
\\
\le{}&
\frac{
\lambda^2(\zeta+1)e^{-\alpha_{\mathrm W}\underline{x}_1}
}{
8\pi^2ND_z^2
}
\sum_{m=1}^{M}
\left[
1+
\frac{
(y_m^{(\mathrm{WG})}-y_k^{(\mathrm{AE})})^2
}{D_z^2}
\right]^{-\frac{\zeta+2}{2}},\nonumber
\end{align}
where the inequality follows from
$r_{k,m_n}^2\ge D_z^2+
(y_m^{(\mathrm{WG})}-y_k^{(\mathrm{AE})})^2$
and
$x_{m_n}^{(\mathrm{PA})}\ge
\underline{x}_n\ge\underline{x}_1$.
Combining this uniform section-norm bound
with \eqref{eq:h_k_norm} and
\eqref{eq:inter_section_correlation_bound},
and noting that $\cos\theta_{k,nn'}\ge0$,
directly yields
\eqref{eq:effective_channel_geometry_bound}.
\end{proof}

Combining Theorem~\ref{thm:coupling_aware_channel_bound}
with the transmit-power normalization in \eqref{e:s},
the achievable rate is upper-bounded by
\begin{align}
\label{eq:rate_geometry_upper_bound}
R_k
\!\le\!
\log_2\!\Bigg[
\!1+&
\frac{
P\lambda^2(\zeta\!+\!1)e^{-\alpha_{\mathrm W}\underline{x}_1}
}{
8\pi^2\sigma^2D_z^2
}
\!\Big(
\!1\!+\!\frac{2}{N}\!\!
\sum_{1\le n<n'\le N}
\!\!\!\!\!\cos\theta_{k,nn'}
\!\!\Big)
\nonumber\\
&\times
\sum_{m=1}^{M}
\Big(
1+
\frac{
(y_m^{(\mathrm{WG})}-y_k^{(\mathrm{AE})})^2
}{D_z^2}
\Big)^{-\frac{\zeta+2}{2}}
\Bigg].
\end{align}
The inter-section coherence factor
$1+\frac{2}{N}\sum_{n<n'}\cos\theta_{k,nn'}$
lies in $[1,N]$. 
Its upper limit recovers the fully coherent-combining ceiling.
Together with guided-wave attenuation and transverse deployment geometry, this factor determines
the rate ceiling. We next investigate its scaling with the number of waveguides under single-agent transmission.

\begin{lem}[Multi-Waveguide Single-Agent Rate-Ceiling Scaling]
\label{lem:rate_bound_scaling}
Consider $M$ equally spaced waveguides with adjacent
spacing $d$, jointly serving a single physical agent
at an arbitrary transverse position $y^{(\mathrm{AE})}$,
measured from the array center.
Under the continuum approximation,
the rate ceiling in \eqref{eq:rate_geometry_upper_bound}
is approximated by
\begin{align}
\label{eq:rate_bound_scaling_M}
&\overline R(M)
\approx\log_2\Bigg[1+
\frac{
P\lambda^2e^{-\alpha_{\mathrm W}\underline{x}_1}
}{
2\pi^2\sigma^2D_zd
}
\mathcal A(\{\theta_{nn'}\})
\\
&\times\Bigg(
\mathcal F\left(
\frac{Md/2-y^{(\mathrm{AE})}}{D_z}
\right)
+
\mathcal F\left(
\frac{Md/2+y^{(\mathrm{AE})}}{D_z}
\right)
\Bigg)\Bigg],\nonumber
\end{align}
where
$\mathcal F(u)\triangleq
\frac{u(3+2u^2)}{[3(1+u^2)^{3/2}]}$.For the small- and large-aperture regimes,
the corresponding asymptotic expressions are
\begin{equation}
\label{eq:rate_bound_asymptotic}
\raisebox{-0.5\height}{%
\resizebox{0.88\columnwidth}{!}{$
\displaystyle
\overline R(M)\approx
\begin{cases}
\displaystyle
\log_2\Bigg[
1+
\frac{
P\lambda^2e^{-\alpha_{\mathrm W}\underline{x}_1}
\mathcal A(\{\theta_{nn'}\})M
}{
2\pi^2\sigma^2D_z^2
\left[1+\left(
\frac{y^{(\mathrm{AE})}}{D_z}
\right)^2\right]^{\frac{5}{2}}
}
\Bigg],
& Md\ll D_z,
\\[4mm]
\displaystyle
\log_2\Bigg[
1+
\frac{
2P\lambda^2e^{-\alpha_{\mathrm W}\underline{x}_1}
\mathcal A(\{\theta_{nn'}\})
}{
3\pi^2\sigma^2D_zd
}
\Bigg],
&
\frac{Md}{2}-|y^{(\mathrm{AE})}|\gg D_z.
\end{cases}
$}}
\end{equation}
where
$\mathcal A(\{\theta_{nn'}\})\triangleq
1+\frac{2}{N}\sum_{n<n'}\cos\theta_{nn'}$
denotes the inter-section coherence factor,
with $\theta_{nn'}$ being the corresponding
inter-section angular bound.
\end{lem}

\begin{proof}
Please see the supplementary material.
\end{proof}

Lemma~\ref{lem:rate_bound_scaling} shows that the transverse
geometric contribution grows approximately linearly with
$M$ for a small aperture, with a reduced gain as the agent
moves away from the array center, and approaches a finite
limit for a sufficiently large aperture.
The overall rate ceiling also depends on guided-wave
attenuation and inter-section coherence.
We next use the general multi-agent rate bound \cite{shao2024theory} to establish
a necessary condition on the number of waveguides for
satisfying prescribed minimum-rate requirements.

\begin{cor}[Necessary Waveguide Count]
\label{cor:waveguide_number_range}
For $M$ equally spaced waveguides with adjacent
spacing $d\ge a$, satisfying $R_k\ge R_0$ for all
$K$ physical agents requires
\begin{align}
\label{eq:waveguide_number_range}
&\Bigg\lceil
\frac{
8\pi^2\sigma^2D_z^2(2^{R_0}-1)
e^{\alpha_{\mathrm W}\underline{x}_1}
}{
P\lambda^2(\zeta+1)
}
\sum_{k=1}^{K}
\left(
\!1\!+\!\frac{2}{N}
\sum_{n<n'}\cos\theta_{k,nn'}
\!\right)^{-1}
\Bigg\rceil
\nonumber\\
&\qquad\le M
\le 1+\left\lfloor\frac{D_y}{d}\right\rfloor .
\end{align}
\end{cor}

\begin{proof}
Please see the supplementary material.
\end{proof}

Corollary~\ref{cor:waveguide_number_range}
establishes a necessary waveguide-count range
governed by the rate target, guided-wave attenuation,
and inter-section coherence, with the upper limit
set by the available transverse space.
Since $\theta_{k,nn'}$ depends on waveguide and
PA placement, the lower bound provides a
deployment-dependent feasibility condition
rather than an explicit minimum waveguide count.

\section{Sum-Rate Maximization}\label{sec:V}
\label{subsec:joint_optimization}

We formulate a sum-rate maximization problem with
minimum-rate guarantees by jointly optimizing the
waveguide positions $\{y_m^{(\mathrm{WG})}\}$,
PA positions $\{x_{m_n}^{(\mathrm{PA})}\}$,
and digital precoder $\bm W$.
For the considered physical-agent locations and the transverse
deployment interval, the problem is formulated as
\begin{subequations}\label{prob:sumrate}
\begin{align}
\text{(P1):}\quad
&\max_{\substack{
\{x_{m_n}^{(\mathrm{PA})}\},\,
\{y_m^{(\mathrm{WG})}\},\,\bm W
}}
\quad
R_{\mathrm{sum}}
=
\sum_{k=1}^{K}R_k,
\label{e:P1a}
\\
\text{s.t.}\quad
&\operatorname{tr}\!\left(\bm W\bm W^{\mathsf H}\right)\le1,
\label{e:P1b}
\\
&0<x_{m_n}^{(\mathrm{PA})}\le D_x,
\label{e:P1c}
\\
&x_{m_{n+1}}^{(\mathrm{PA})}
-x_{m_n}^{(\mathrm{PA})}
\ge\frac{\lambda}{2},
\quad n<N,
\label{e:P1d}
\\
&0\le y_m^{(\mathrm{WG})}\le D_y,
\label{e:P1e}
\\
&d_{mm'}^{(\mathrm{WG})}\ge a,
\quad m\ne m',
\label{e:P1f}
\\
&\left\|
\bm\psi_k^{(\mathrm{AE})}
-\bm\psi_{k'}^{(\mathrm{AE})}
\right\|
\ge d_{kk'}^{(\mathrm{AE})},
\quad k\ne k',
\label{e:P1g}
\\
&R_k\ge R_0,
\quad k=1,\ldots,K.
\label{e:P1h}
\end{align}
\end{subequations}
Here, $R_0$ is the minimum rate required by each physical agent,
and $R_k$ is the achievable rate defined in \eqref{eq:k_rate}.

Constraint \eqref{e:P1b} normalizes the transmit power, while
\eqref{e:P1c}--\eqref{e:P1d} constrain the PA positions and spacing,
and \eqref{e:P1e}--\eqref{e:P1f} constrain the waveguide deployment
and prevent core overlap.
Constraint \eqref{e:P1g} specifies the minimum separation
between physical agents, while \eqref{e:P1h} imposes the
minimum-rate requirement.

Problem (P1) is nonconvex due to multi-agent interference and the
nonlinear dependence of the effective channel on the waveguide
and PA positions.
In the following sections, we develop a WMMSE-SCA-based
alternating optimization algorithm and a signal-domain PDD
algorithm to solve (P1), both using the full end-to-end
channel in \eqref{eq:end_to_end}.

\subsection{WMMSE-Based Alternating Optimization}
\label{subsubsec:wmmse_ao}

We first transform the sum-rate maximization problem into a weighted
minimum mean-square error (WMMSE) problem. Let $u_k\in\mathbb C$
denote the scalar receiver used by physical agent $k$ to estimate
$s_{0,k}$ from $\xi_k$. According to \eqref{eq:user_received}, the
corresponding mean-square error (MSE) is \cite{shi2011wmmse}
\begin{equation}
e_k
\!=\!
\left|
1-u_k^*\sqrt{P}\bm h_k^{\mathsf H}\bm w_k
\right|^2
\!\!+\!
P\sum_{i\ne k}
\left|
u_k^*\bm h_k^{\mathsf H}\bm w_i
\right|^2
\!+\!\sigma^2|u_k|^2 .
\label{eq:wmmse_mse}
\end{equation}
Introducing $u_k\in\mathbb C$ and $v_k>0$, the achievable
rate admits the WMMSE representation
$R_k=\max_{u_k,v_k}\frac{1+\ln v_k-v_ke_k}{\ln 2}.$
Thus, (P1) is equivalently reformulated as
\begin{subequations}
\label{prob:wmmse}
\begin{align}
\min_{\substack{
\{x_{m_n}^{(\mathrm{PA})}\},
\{y_m^{(\mathrm{WG})}\},
\bm W,\{u_k,v_k\}
}}
\quad&
\sum_{k=1}^{K}
\left(v_ke_k-\ln v_k\right)
\label{eq:wmmse_obj}
\\
\mathrm{s.t.}\quad&
\eqref{e:P1b}-\eqref{e:P1f},
\label{eq:wmmse_geometry_constraints}
\\
& \hspace{-4cm}
v_ke_k-\ln v_k
\le
1-R_0\ln2,
\quad k=1,\ldots,K .
\label{eq:wmmse_qos}
\end{align}
\end{subequations}
Here, \eqref{eq:wmmse_qos} is the WMMSE form of the minimum-rate
constraint in \eqref{e:P1h}. The physical-agent separation constraint
\eqref{e:P1g} does not participate in the optimization since the
physical-agent locations are fixed.
We solve \eqref{prob:wmmse} by alternating optimization over
$\{u_k\}$, $\{v_k\}$, $\bm W$, the PA positions
$\{x_{m_n}^{(\mathrm{PA})}\}$, and the waveguide positions
$\{y_m^{(\mathrm{WG})}\}$. For fixed $\bm W$ and spatial deployment,
the receiver and weight updates admit the closed-form solutions
\begin{equation} 
\label{eq:wmmse_uv_update}
u_k^\star
=
\frac{
\sqrt{P}\bm h_k^{\mathsf H}\bm w_k
}{
P\sum_{i=1}^{K}
|\bm h_k^{\mathsf H}\bm w_i|^2+\sigma^2
},\quad
v_k^\star
=
\left(e_k^\star\right)^{-1}.
\end{equation}

With $\{u_k,v_k\}$ and the spatial deployment fixed, the precoder
is obtained from
\begin{subequations}
\label{prob:wmmse_precoder}
\begin{align}
\min_{\bm W}\quad&
\sum_{k=1}^{K}v_ke_k
\\
\mathrm{s.t.}\quad&
\operatorname{tr}(\bm W\bm W^{\mathsf H})\le1,
\\
&
v_ke_k-\ln v_k
\le1-R_0\ln2,
\quad \forall k .
\end{align}
\end{subequations}
Since $e_k$ is a convex quadratic function of $\bm W$ for fixed
channel and WMMSE variables, \eqref{prob:wmmse_precoder} is a
convex quadratically constrained quadratic program.

We next optimize the spatial deployment. Let $\bm x$ collect all
PA longitudinal positions and let $\bm y$ collect all waveguide
transverse positions. For fixed $\bm W$, $\{u_k,v_k\}$, and the
other geometry block, the MSE remains nonconvex with respect to
either $\bm x$ or $\bm y$ because the spatial variables enter the
coupled waveguide propagation and the PA-to-agent channel.
Accordingly, successive convex approximation (SCA) is applied.
For either geometry block
$\bm z\in\{\bm x,\bm y\}$, at the current point $\bm z^{(t)}$,
we construct
\begin{equation}
\widehat e_k(\bm z)
\!=\!
e_k(\bm z^{(t)})
\!+\!
\nabla_{\bm z}e_k(\bm z^{(t)})^{\top}
\!(\bm z\!-\!\bm z^{(t)})
\!+\!
\frac{\rho_z}{2}
\|\bm z\!-\!\bm z^{(t)}\|_2^2 ,
\label{eq:wmmse_sca}
\end{equation}
where $\rho_z>0$ is selected to make
$\widehat e_k(\bm z)$ a local convex upper approximation of
$e_k(\bm z)$.

The PA-position block is then updated by minimizing
$\sum_kv_k\widehat e_k(\bm x)$ subject to
\eqref{e:P1c}-\eqref{e:P1d} and
$v_k\widehat e_k(\bm x)-\ln v_k\le1-R_0\ln2$.
Similarly, the waveguide-position block minimizes
$\sum_kv_k\widehat e_k(\bm y)$ subject to
\eqref{e:P1e}-\eqref{e:P1f} and the corresponding WMMSE rate
constraints. 
The algorithm therefore updates
$\{u_k\}$, $\{v_k\}$, $\bm W$,
$\{x_{m_n}^{(\mathrm{PA})}\}$, and
$\{y_m^{(\mathrm{WG})}\}$ sequentially in each iteration.
The full channel in \eqref{eq:end_to_end} is recomputed after each
spatial update, and the procedure terminates when the improvement
in the sum rate falls below a prescribed threshold.

\begin{figure*}[ht]
\centering
\begin{subfigure}[t]{0.25\textwidth}
    \centering
    \includegraphics[width=\linewidth]{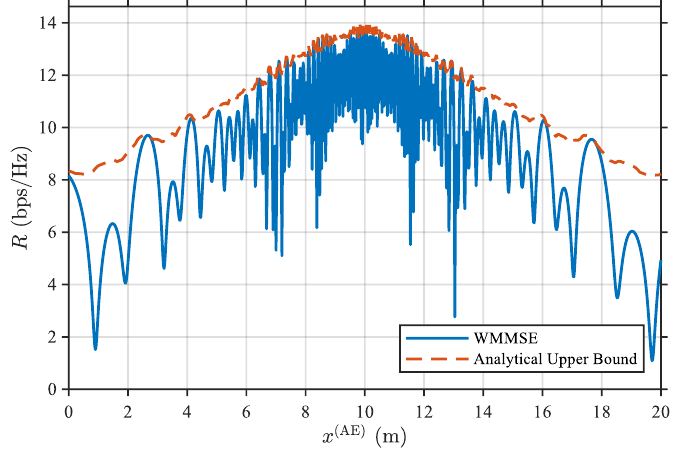}
    \caption{}
    \label{fig:rate_x_cut}
\end{subfigure}
\hfill
\begin{subfigure}[t]{0.25\textwidth}
    \centering
    \includegraphics[width=\linewidth]{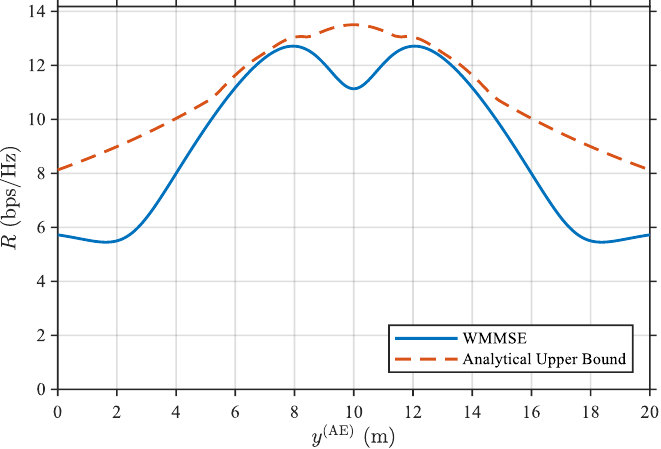}
    \caption{}
    \label{fig:rate_y_cut}
\end{subfigure}
\hfill
\begin{subfigure}[t]{0.25\textwidth}
    \centering
    \includegraphics[width=\linewidth]{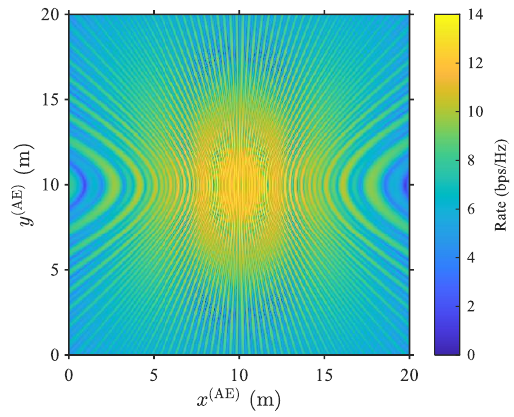}
    \caption{}
    \label{fig:rate_spatial_map}
\end{subfigure}
\caption{Spatial rate characteristics of the two-waveguide
single-agent CWPASS:
(a) achievable rate versus $x^{(\mathrm{AE})}$ at
$y^{(\mathrm{AE})}=10$ m;
(b) achievable rate versus $y^{(\mathrm{AE})}$ at
$x^{(\mathrm{AE})}=10$ m;
(c) spatial distribution of the WMMSE achievable rate.}
\label{fig:spatial_rate}
\end{figure*}

\subsection{Signal-Domain PDD}
\label{subsubsec:signal_pdd}

Unlike WMMSE-SCA, which updates PA and waveguide
positions separately, signal-domain PDD enables
a joint spatial update.
We introduce $\tilde{\bm Y}\in\mathbb C^{K\times K}$
subject to
\begin{equation}
\tilde{\bm Y}
=
\sqrt{P}\bm H\bm W .
\label{eq:pdd_consistency}
\end{equation}
The MSE in \eqref{eq:wmmse_mse} becomes
\begin{equation}
e_k
=
\left|1-u_k^*\tilde Y_{kk}\right|^2
+
\sum_{i\ne k}
\left|u_k^*\tilde Y_{ki}\right|^2
+
\sigma^2|u_k|^2 .
\label{eq:pdd_mse}
\end{equation}
Thus, the WMMSE objective and QoS constraints
depend on $\{\bm W,\bm x,\bm y\}$ only through
\eqref{eq:pdd_consistency}.

Using PDD \cite{shi2020pdd}, with dual matrix
$\bm\Lambda\in\mathbb C^{K\times K}$ and penalty
parameter $\rho>0$, the augmented-Lagrangian
objective is
\begin{equation}
\mathcal L_{\rho}
=
\sum_{k=1}^{K}
\left(v_ke_k-\ln v_k\right)
+
\frac{\rho}{2}
\left\|
\tilde{\bm Y}
-\sqrt{P}\bm H\bm W
+\frac{\bm\Lambda}{\rho}
\right\|_{\mathrm F}^{2},
\label{eq:pdd_augmented_objective}
\end{equation}
subject to the constraints in \eqref{prob:wmmse},
with \eqref{eq:wmmse_qos} evaluated using
\eqref{eq:pdd_mse}.
The inner problem alternately updates
$\{u_k,v_k\}$, $\tilde{\bm Y}$, $\bm W$,
and the joint deployment $(\bm x,\bm y)$.

For fixed $\tilde{\bm Y}$, the receiver and weight
updates follow \eqref{eq:wmmse_uv_update},
replacing $\sqrt{P}\bm h_k^{\mathsf H}\bm w_i$
with $\tilde Y_{ki}$.
The $\tilde{\bm Y}$ subproblem is separable
across rows, with unconstrained minimizer
\begin{equation}
[\tilde{\bm Y}]_{k,:}^{\mathrm{unc}}
=
\frac{
2v_ku_k[\bm I_K]_{k,:}
+
\rho\sqrt{P}[\bm H\bm W]_{k,:}
-
[\bm\Lambda]_{k,:}
}{
2v_k|u_k|^2+\rho
}.
\label{eq:pdd_ytilde_update}
\end{equation}
If the QoS constraint is violated, the solution
is projected onto the feasible MSE ball centered
at $[\bm I_K]_{k,:}/u_k^*$ with radius
$\sqrt{(1+\ln v_k-R_0\ln2)/(v_k|u_k|^2)-\sigma^2}$,
obtained by completing the square in
\eqref{eq:pdd_mse}.

With $\tilde{\bm Y}$ and the spatial deployment fixed,
the precoder update reduces to constrained
least squares:
\begin{equation}
\bm W
=
\left(
P\bm H^{\mathsf H}\bm H
+\varrho\bm I_M
\right)^{\dagger}
\sqrt{P}\bm H^{\mathsf H}
\left(
\tilde{\bm Y}
+\frac{\bm\Lambda}{\rho}
\right),
\label{eq:pdd_precoder_update}
\end{equation}
where $(\cdot)^\dagger$ denotes the Moore--Penrose
pseudoinverse, and $\varrho\ge0$ enforces the
transmit-power constraint and is obtained by
bisection when the constraint is active.

To jointly update the PA and waveguide positions,
let $\bm J_x$ and $\bm J_y$ denote the Jacobians of
$\operatorname{vec}(\sqrt{P}\bm H\bm W)$ with respect
to $\bm x$ and $\bm y$, respectively.
Linearizing around the current deployment gives
\begin{equation}
\begin{aligned}
\min_{\Delta\bm x,\Delta\bm y}\quad&
\Bigg\|
\operatorname{vec}\!\left(
\tilde{\bm Y}
-\sqrt{P}\bm H\bm W
+\frac{\bm\Lambda}{\rho}
\right)
-\bm J_x\Delta\bm x
-\bm J_y\Delta\bm y
\Bigg\|_2^2
\\
&+
\chi_x\|\Delta\bm x\|_2^2
+\chi_y\|\Delta\bm y\|_2^2
\end{aligned}
\label{eq:pdd_joint_geometry_update}
\end{equation}
subject to \eqref{e:P1c}--\eqref{e:P1f},
where $\chi_x,\chi_y>0$ are regularization parameters.
The full channel in \eqref{eq:end_to_end}
is recomputed after each spatial update.
After solving the inner problem, the consistency
error in \eqref{eq:pdd_consistency} is evaluated.
If sufficiently small, the dual matrix is updated as
$\bm\Lambda\leftarrow
\bm\Lambda+
\rho(\tilde{\bm Y}-\sqrt{P}\bm H\bm W)$;
otherwise, $\rho$ is increased.
The iterations continue until the consistency
error and objective variation satisfy the
stopping criteria.

\section{Numerical Results}\label{sec:VI}

We consider a service region of $20\times20\times3~\mathrm{m}^3$ at $28~\mathrm{GHz}$.
A compact coupled-waveguide array with $M=6$ dielectric waveguides is deployed, with each waveguide carrying $N=3$ PAs.
The dielectric waveguides have a rectangular cross section of $6.0\times4.0~\mathrm{mm}^2$, with refractive indices
$n_1=1.5$ and $n_0=1$ for the core and cladding, respectively.
The total input power and noise power are set to $20~\mathrm{dBm}$ and $-80~\mathrm{dBm}$, respectively, while the in-waveguide loss is $0.08~\mathrm{dB/m}$.
The cosine-power radiation exponent is set to $\zeta=3$, and the minimum-rate requirement is $R_0=1~\mathrm{bit/s/Hz}$. 
A total of $K=6$ physical agents are randomly generated within two circular hotspots of radius $1.5~\mathrm{m}$, centered at $(6,4,0)$ and $(14,6,0)$, respectively, with three physical agents in each hotspot and the minimum separation $d_{kk'}^{(\mathrm{AE})}=0.5~\mathrm{m}$.

\subsection{Single-Agent Spatial Rate Analysis}
\label{subsec:spatial_rate}

We first examine the spatial rate characteristics of CWPASS
and the tightness of the analytical upper bound.
Consider two coupled waveguides with $N=3$ PAs each,
centered at $y=10$ m with spacing $d=12$ mm,
and PA sections at $x=9.7$, $10$, and $10.3$ m.
A single physical agent moves across the service region
while the waveguide and PA positions remain fixed.
Fig.~\ref{fig:spatial_rate} compares the achievable rate
with the analytical upper bound.
Fig.~\ref{fig:rate_x_cut} reveals pronounced rate
oscillations caused by constructive and destructive
combining among distributed PAs, while the upper bound
captures the potential coherent gain.
Fig.~\ref{fig:rate_y_cut} exhibits an approximately
symmetric profile with two off-center peaks,
indicating that shorter propagation distances
do not necessarily yield higher rates.
The spatial rate map in Fig.~\ref{fig:rate_spatial_map}
further reveals dense peaks and nulls, highlighting
the sensitivity of coherent combining to agent location.
These results demonstrate the joint influence of
propagation geometry and inter-section phase relationships
on spatial rate performance.

\begin{figure}[!tb]
    \centering
    \includegraphics[width=0.76\linewidth]{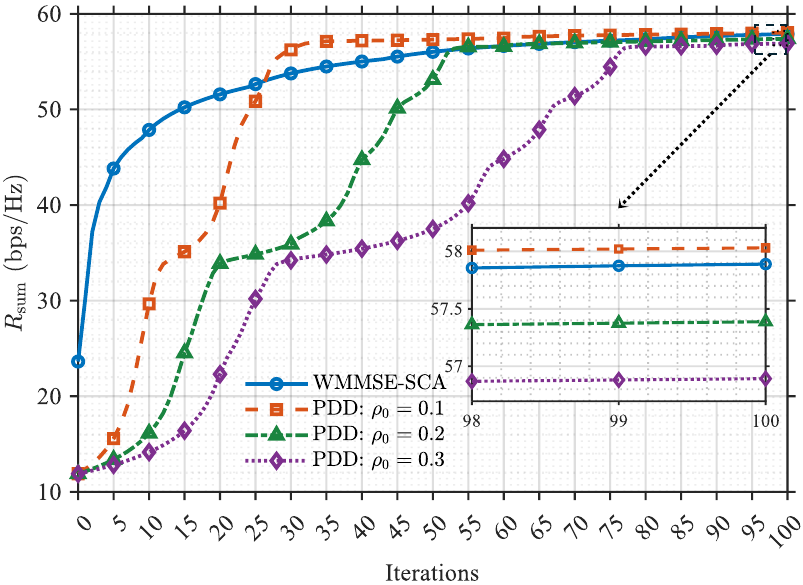}
    \caption{Sum-rate convergence of the proposed WMMSE and PDD algorithms under different initial penalty parameters.}
    \label{fig:convergence}
\end{figure}

\subsection{Algorithm Performance}

We first evaluate the convergence behavior of the proposed
WMMSE-SCA and PDD algorithms, followed by a comparison of
their achievable rates under different transmit power levels.

\subsubsection{Convergence Behavior}

Fig.~\ref{fig:convergence} compares the sum-rate
convergence of WMMSE-SCA and PDD under different
initial penalty parameters $\rho_0$.
The main observations are as follows.

\begin{itemize}
    \item \textit{Convergence of WMMSE-SCA:}
    WMMSE-SCA exhibits rapid initial improvement,
    reaching approximately 50~bps/Hz within
    15 iterations, followed by gradual convergence
    toward 58~bps/Hz.

    \item \textit{Impact of the initial penalty parameter:}
    PDD convergence is sensitive to $\rho_0$.
    With $\rho_0=0.1$, PDD converges fastest,
    approaching 58~bps/Hz within 35 iterations.
    Larger values of $\rho_0$ lead to slower
    convergence and slightly lower final rates,
    as highlighted by the inset.

    \item \textit{Comparison of the proposed algorithms:}
    WMMSE-SCA achieves faster initial convergence,
    while PDD with $\rho_0=0.1$ reaches a
    comparable final rate with fewer iterations.
    Overall, both algorithms achieve similar
    sum-rate performance, although appropriate
    penalty initialization is important for PDD.
\end{itemize}

\subsubsection{Achievable Rate}

\begin{figure}[!tb]
    \centering
    \includegraphics[width=0.76\linewidth]{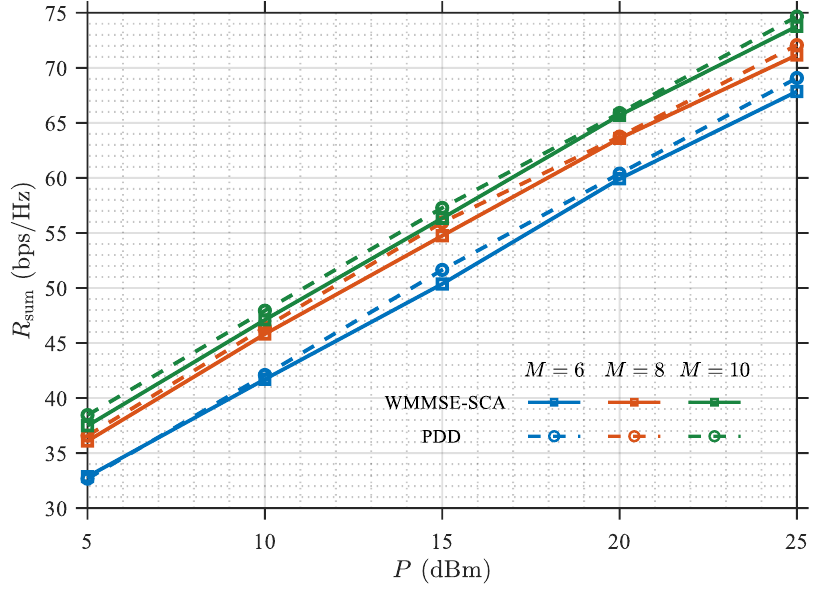}
    \caption{Sum-rate of the proposed algorithms versus transmit power for different numbers of coupled waveguides.}
    \label{fig:algo_rate}
\end{figure}

\begin{figure}[!tb]
    \centering
    \includegraphics[width=0.76\linewidth]{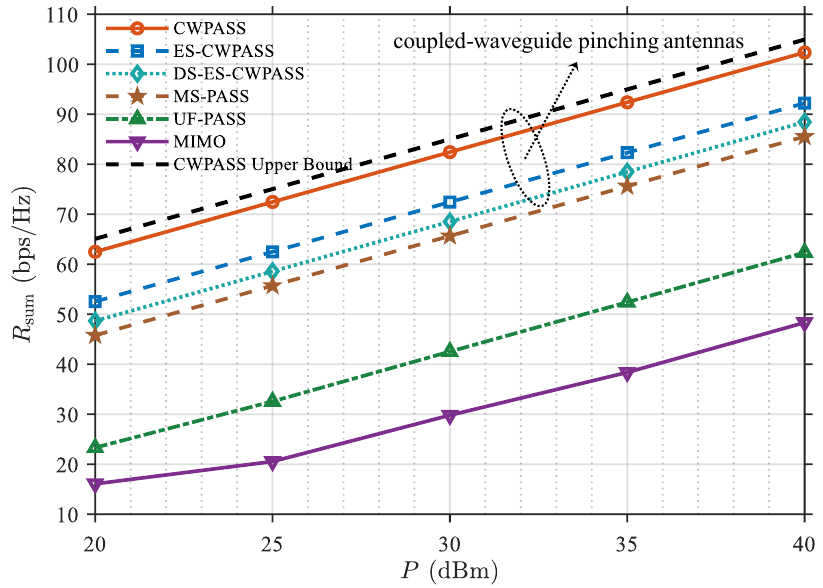}
    \caption{Achievable sum rate versus transmit power for
    different architectures.}
    \label{fig:r_t}
\end{figure}

Fig.~\ref{fig:algo_rate} compares the sum-rate performance
of the proposed PDD and WMMSE-SCA algorithms under
different transmit power levels and numbers of coupled
waveguides, $M=6,8,10$.
The achievable sum rate increases with both transmit power
and the number of waveguides, as additional waveguides
provide more RF chains and radiating PAs, thereby
increasing the available spatial degrees of freedom.
Both algorithms achieve comparable performance, with
PDD generally yielding slightly higher sum rates
in most of the considered scenarios.
Unless otherwise specified, PDD is employed for the
remaining numerical evaluations.

\subsection{CWPASS Performance}

We now evaluate the performance of the proposed CWPASS framework, focusing on the advantages offered by inter-waveguide coupling and spatial reconfigurability. 
To assess these advantages, we consider the following five system configurations:
\begin{itemize}[leftmargin = 0.5cm]
    \item \textbf{CWPASS:} The proposed coupled-waveguide PASS,
    where the waveguide positions and PA locations are jointly
    optimized to exploit inter-waveguide coupling.

    \item \textbf{ES-CWPASS:} Equally spaced CWPASS, where
    adjacent waveguides maintain identical spacing, while
    the common spacing, array center, and PA locations
    remain continuously optimizable.

    \item \textbf{DS-ES-CWPASS:} Discrete-spacing equally
    spaced CWPASS, where the common waveguide spacing is selected
    from the feasible candidates motivated by the
    two-waveguide phase-matching rule in
    Corollary~\ref{cor:optimal_waveguide_spacing},
    with $p\in\{1,\ldots,20\}$ under a fixed reference
    geometry. For each candidate, the array center,
    PA locations, and digital precoder are jointly optimized.

    \item \textbf{MS-PASS:} Minimum-spacing PASS, where
     the waveguide positions and PA locations are jointly
     optimized subject to a minimum adjacent-waveguide
     spacing of $D_{\mathrm{min}}^{(\mathrm{WG})}=0.5$ m \cite{ding2026movable}, effectively suppressing
     inter-waveguide coupling.

    \item \textbf{UF-PASS:} Conventional PASS with waveguides
    uniformly distributed along the $y$-direction of the
    service region. The waveguide positions are fixed,
    whereas the PA locations remain adjustable.

    \item \textbf{MIMO:} A conventional hybrid MIMO system
    with a BS located at $(D_x/2,0,D_z)$, equipped with
    $MN$ antenna elements arranged in a half-wavelength-spaced
    planar array in the $x$-$z$ plane and $M$ RF chains.
\end{itemize}

\subsubsection{Sum Rate vs. Transmit Power}

Fig.~\ref{fig:r_t} presents the achievable sum rate versus
transmit power for the proposed CWPASS and benchmark schemes.
We have the following main observations.

\begin{figure}[!tb]
    \centering
    \includegraphics[width=0.76\linewidth]{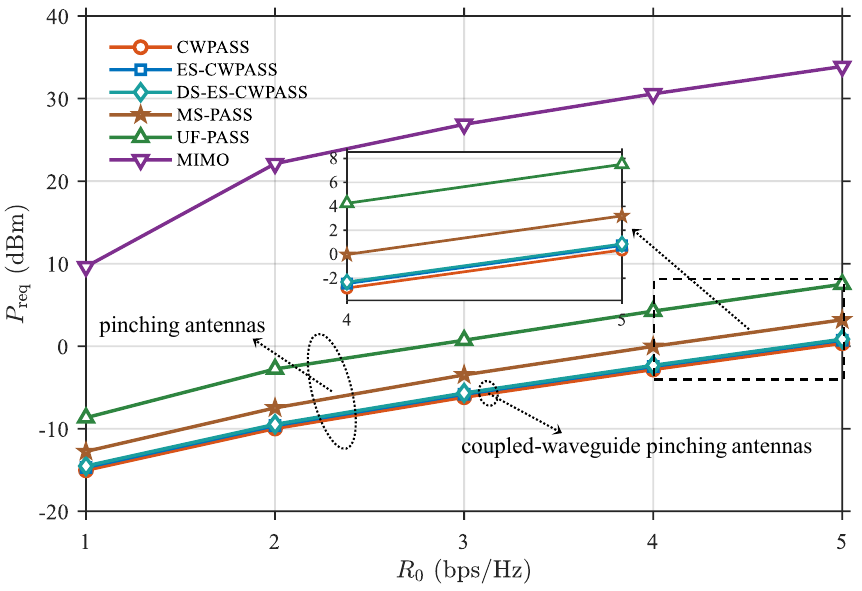}
    \caption{Required transmit power versus per-agent QoS
    target for different architectures.}
    \label{fig:qos_power}
\end{figure}

\begin{itemize}[leftmargin=0.4cm]

\item The sum rates of all schemes increase with
transmit power, with CWPASS consistently achieving
the highest performance.
At $P=40$ dBm, CWPASS achieves approximately
$102$ bps/Hz, compared with $92$, $89$, $86$,
$62$, and $48$ bps/Hz for ES-CWPASS,
DS-ES-CWPASS, MS-PASS, UF-PASS, and MIMO,
respectively.
Moreover, CWPASS closely approaches its
deployment-dependent upper bound of approximately
$105$ bps/Hz.

\item CWPASS outperforms ES-CWPASS by approximately
$10$ bps/Hz, while ES-CWPASS further outperforms
DS-ES-CWPASS by $3$ bps/Hz.
This demonstrates the benefits of nonuniform
waveguide deployment and continuous spacing
optimization.
Nevertheless, DS-ES-CWPASS achieves competitive
performance using the finite phase-matching-based
spacing candidates derived from
Corollary~\ref{cor:optimal_waveguide_spacing}.

\item MS-PASS outperforms UF-PASS by approximately
$24$ bps/Hz.
Since inter-waveguide coupling is negligible
in both schemes, this gain primarily stems from
waveguide position optimization, which enables
the radiating PAs to approach clustered AEs
and reduce propagation loss.

\item CWPASS further outperforms MS-PASS by
approximately $16$ bps/Hz.
Unlike MS-PASS, CWPASS allows closely spaced
waveguides, enabling compact deployment and
exploitable evanescent-field coupling.
This demonstrates that inter-waveguide coupling,
rather than being merely an unwanted interaction,
can provide an additional degree of freedom
for coherent signal combining and wireless
channel reconfiguration.

\item All PASS schemes outperform conventional
MIMO, with even UF-PASS achieving a gain of
approximately $14$ bps/Hz, while CWPASS
provides a gain of $54$ bps/Hz.
This highlights the advantages of distributed
and adjustable PA radiation over fixed
antenna arrays, particularly in multi-hotspot
communication scenarios.
\end{itemize}

\subsubsection{Required Transmit Power vs. QoS Target}

Fig.~\ref{fig:qos_power} illustrates the required transmit
power versus the per-agent QoS target $R_0$, with
waveguide deployment, PA placement, and digital precoding
reoptimized for each target.
The required power increases with $R_0$ for all schemes,
with CWPASS consistently achieving the lowest power
consumption.
As highlighted in the inset, ES-CWPASS and DS-ES-CWPASS
closely approach CWPASS, whereas MS-PASS requires
higher power but still outperforms UF-PASS.
This demonstrates the benefits of flexible waveguide
deployment even under minimum-spacing constraints.
Moreover, the substantial power advantage over
conventional MIMO highlights the effectiveness of
distributed PA radiation, while the additional gains
of CWPASS reflect the potential of compact deployment
and inter-waveguide coupling.

\subsubsection{Sum Rate vs. Number of Physical Agents}

\begin{figure}[!tb]
    \centering
    \begin{subfigure}{\linewidth}
        \centering
        \includegraphics[width=0.72\linewidth]{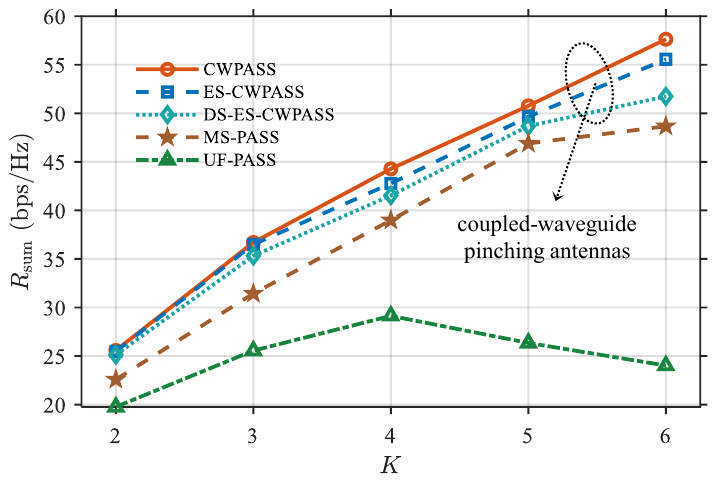}
        \caption{}\label{fig:rate_K}
    \end{subfigure}

    \vspace{-0.1cm}

    \begin{subfigure}{\linewidth}
        \centering
        \includegraphics[width=0.72\linewidth]{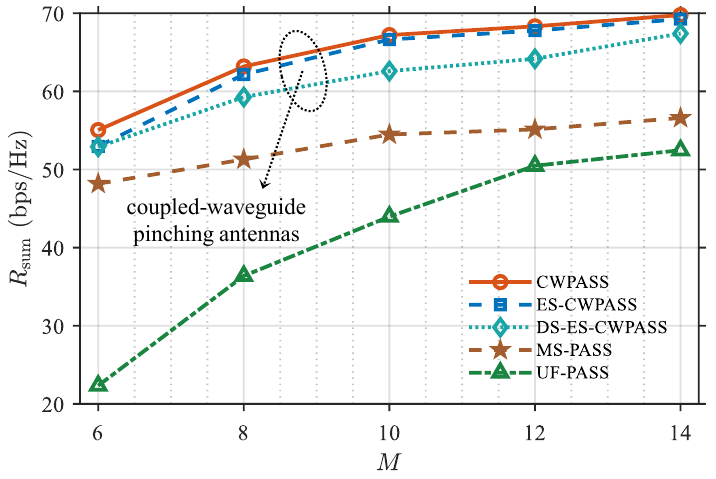}
        \caption{}\label{fig:rate_M}
    \end{subfigure}
    \caption{Achievable sum rate versus (a) the number
    of physical agents $K$ and (b) the number of
    waveguides $M$ for different PASS schemes.}
    \label{fig:rate_K_M}
\end{figure}

Fig.~\ref{fig:rate_K} illustrates the achievable sum rate
versus the number of physical agents $K$.
CWPASS consistently outperforms all benchmarks,
with its sum rate continuing to increase as more
agents are served.
While ES-CWPASS and DS-ES-CWPASS exhibit similar
increasing trends, the improvement of MS-PASS
gradually diminishes, whereas UF-PASS even suffers
a rate degradation at larger $K$.
These results indicate that flexible waveguide
deployment becomes increasingly important for
supporting additional agents under a fixed
transmit-power budget.
Moreover, the performance gap between CWPASS
and MS-PASS highlights the benefits of allowing
closely spaced waveguides, which enables more
compact deployment and potentially exploitable
inter-waveguide coupling.

\subsubsection{Sum Rate vs. Number of Waveguides}

Fig.~\ref{fig:rate_M} shows that the achievable
sum rates generally increase with $M$, as
additional waveguides provide more spatial
degrees of freedom for precoding and PA deployment.
CWPASS consistently achieves the highest sum rate,
while its performance gap with ES-CWPASS narrows
as $M$ increases, indicating that optimized
equal spacing becomes increasingly effective.
Moreover, the coupled-waveguide schemes
substantially outperform MS-PASS and UF-PASS,
highlighting the additional gains enabled by
exploiting inter-waveguide coupling.
The diminishing rate improvements at larger
$M$ also suggest that the benefits of adding
waveguides gradually saturate.

\section{Conclusion}

This paper turns inter-waveguide evanescent coupling, conventionally regarded as undesirable crosstalk, into an exploitable beamforming resource for PASS. Beyond the resulting communication gains, this finding challenges the conventional view that waveguides should be electromagnetically isolated to ensure efficient signal delivery. Instead, it suggests that the waveguide infrastructure itself can participate in wireless channel shaping, opening a new dimension for communication system design beyond conventional antenna placement and digital precoding. Importantly, the findings do not imply that stronger coupling is always preferable, but rather that electromagnetic interactions should be engineered to serve communication objectives. This perspective offers a promising path toward more spectrally efficient wireless architectures through the joint design of guided-wave propagation and free-space radiation.

Realizing this potential in practical systems calls for further investigation into full-wave electromagnetic validation, hardware implementation, and robustness against fabrication imperfections and deployment uncertainties. Extending coupling-aware design to broadband transmission and dynamic wireless environments would further enable adaptive exploitation of electromagnetic interactions. Ultimately, such advances could transform PASS from a spatially reconfigurable antenna architecture into a more deeply integrated electromagnetic communication platform, where the transmission infrastructure itself becomes an active design resource for wireless performance enhancement.

\appendices






\bibliographystyle{IEEEtran}
\bibliography{Ref}

\clearpage
\section*{Supplementary Material}

\begin{proof}[Proof of Proposition~\ref{prop:modal_normalization}]
For the isolated guided mode, Maxwell's equation gives
$\bm{\mathcal H}_m^{(\mathrm{iso})}
=-(j\omega\psi_0)^{-1}
\nabla\times\bm{\mathcal E}_m^{(\mathrm{iso})}$.
Under the dominant-field approximation, the
longitudinal magnetic-field component satisfies
$[\bm{\mathcal H}_m^{(\mathrm{iso})}]_z
\!\approx \!\frac{\beta_g[\bm{\mathcal E}_m^{(\mathrm{iso})}]_y}{(\omega\psi_0)}$.
Consequently, the time-averaged Poynting flux
gives the unit-power condition
\begin{align}
1
=&\frac12\operatorname{Re}
\int_{-\infty}^{\infty}
\int_{-\infty}^{\infty}
\Big(
\bm{\mathcal E}_m^{(\mathrm{iso})}
\times
\bm{\mathcal H}_m^{(\mathrm{iso})*}
\Big)\cdot\bm e_x
\,dy\,dz
\nonumber\\
\approx&
\frac{\beta_g}{2\omega\psi_0}
\int_{-\infty}^{\infty}
\int_{-\infty}^{\infty}
\left\|
\bm{\mathcal E}_m^{(\mathrm{iso})}(x,y,z)
\right\|^2
dy\,dz.
\label{eq:unit_power_condition}
\end{align}

Under the separable Marcatili modal envelope,
the transverse field integral factorizes into
the lateral and vertical contributions.
From \eqref{eq:core_mode_field}, the field
inside the core follows a cosine profile,
while the exterior fields decay exponentially.
The lateral integral, including both cladding
tails, is evaluated as
\begin{align}
&\int_{-a/2}^{a/2}
\cos^2(k_yy)\,dy
+
2\cos^2\left(\frac{k_ya}{2}\right)
\int_0^\infty e^{-2\alpha_y y}\,dy
\nonumber\\
=&\frac{a}{2}
+\frac{\sin(k_ya)}{2k_y}
+\frac{\cos^2\left(\frac{k_ya}{2}\right)}
{\alpha_y}.
\end{align}

Similarly, the vertical cladding field decays
with the transverse decay constant
$\sqrt{(\omega/c)^2(n_1^2-n_0^2)-k_z^2}$,
as determined by the modal dispersion relations.
The corresponding vertical integral is
\begin{align}
&\int_{-b/2}^{b/2}
\cos^2(k_zz)\,dz+
2\cos^2\left(\frac{k_zb}{2}\right)
\nonumber\\
&\times
\int_0^\infty
\exp\left[
-2z\sqrt{
\left(\frac{\omega}{c}\right)^2
(n_1^2-n_0^2)-k_z^2
}
\right]dz
\nonumber\\
=&\frac{b}{2}
+\frac{\sin(k_zb)}{2k_z}
+
\frac{
\cos^2\left(\frac{k_zb}{2}\right)
}{
\sqrt{
\left(\frac{\omega}{c}\right)^2
(n_1^2-n_0^2)-k_z^2
}
}.
\end{align}

Combining the two transverse integrals,
the unit-power condition becomes
\begin{align}
1=&
\frac{\beta_g|\mathcal E_0|^2}
{2\omega\psi_0}
\left[
\frac{a}{2}
+\frac{\sin(k_ya)}{2k_y}
+\frac{\cos^2\left(\frac{k_ya}{2}\right)}
{\alpha_y}
\right]
\nonumber\\
&\times
\Bigg[
\frac{b}{2}
+\frac{\sin(k_zb)}{2k_z}
+
\frac{
\cos^2\left(\frac{k_zb}{2}\right)
}{
\sqrt{
\left(\frac{\omega}{c}\right)^2
(n_1^2-n_0^2)-k_z^2
}
}
\Bigg].
\end{align}
Solving for the positive normalization factor
$\mathcal E_0$ yields
\eqref{eq:modal_normalization_factor}.
\end{proof}

\begin{proof}[Proof of Theorem~\ref{thm:pairwise_evanescent_field}]
In the homogeneous cladding, the vector wave
equation \eqref{eq:vector_wave_equation} reduces
to the vector Helmholtz equation
\begin{equation*}
\nabla^2\bm{\mathcal E}_m^{(\mathrm{ev})}
+k_0^2n_0^2\bm{\mathcal E}_m^{(\mathrm{ev})}
=\bm 0.
\end{equation*}
Under the Marcatili approximation, the evanescent
field inherits the longitudinal dependence
$e^{-j\beta_gx}$ and vertical dependence
$\cos[k_z(z-D_z)]$ of the fundamental guided mode.
Neglecting the weak longitudinal attenuation
in the transverse modal equation yields
\begin{equation}
\frac{\partial^2\bm{\mathcal E}_m^{(\mathrm{ev})}}
{\partial y^2}
-\alpha_y^2\bm{\mathcal E}_m^{(\mathrm{ev})}
=\bm 0,
\end{equation}
where $\alpha_y^2=\beta_g^2+k_z^2-k_0^2n_0^2$.
Since the evanescent field must remain bounded
away from the waveguide, only the decaying
solution is retained. For
$y>y_m^{(\mathrm{WG})}+a/2$, we obtain
\begin{align}
\bm{\mathcal E}_m^{(\mathrm{ev})}(x,y,z)
=&\bm{\mathcal E}_m^{(\mathrm{WG})}
\left(x,y_m^{(\mathrm{WG})}+\frac a2,z\right)
\nonumber\\
&\times
\exp\left[
-\alpha_y\left(
y-y_m^{(\mathrm{WG})}-\frac a2
\right)\right].
\label{eq:evanescent_from_boundary}
\end{align}
From \eqref{eq:core_mode_field} and
\eqref{eq:guided_field_CWPASS}, the field
at the right core boundary is
\begin{align*}
\bm{\mathcal E}_m^{(\mathrm{WG})}
\left(x,y_m^{(\mathrm{WG})}+\frac a2,z\right)
\approx&
\mathcal E_0s_m
e^{-\frac{\alpha_{\mathrm W}}2x}
e^{-j\beta_gx}
\cos\left(\frac{k_ya}{2}\right)
\nonumber\\
&\times\cos[k_z(z-D_z)]\bm e_y.
\end{align*}

For $(y,z)\in\mathcal S_{m'}$, the transverse
decay can be separated into attenuation
across the inter-waveguide gap and penetration
into the core of waveguide $m'$, giving
\begin{align}
&e^{-\alpha_y(y-y_m^{(\mathrm{WG})}-a/2)}
=\exp\left[
-\alpha_y\left(
\frac a2+y-y_{m'}^{(\mathrm{WG})}
\right)\right]
\nonumber\\
&\times
\exp\left[
-\alpha_y\left(
\left|y_m^{(\mathrm{WG})}
-y_{m'}^{(\mathrm{WG})}\right|-a
\right)\right].
\end{align}
Substituting the boundary field and the
above decay decomposition into
\eqref{eq:evanescent_from_boundary}
yields \eqref{eq:evanescent_field_mm}.
\end{proof}

\begin{proof}[Proof of Corollary~\ref{cor:optimal_waveguide_spacing}]
For $M=2$, $\operatorname{diag}(\bm C)=\bm 0$ and
$[\bm C]_{12}=[\bm C]_{21}=\mu(d)$.
Using the Taylor expansion of the matrix exponential, we obtain
\begin{align}
&e^{j\bm C(L_{n'}\!-\!L_n)} \nonumber \\
&\!=\!
\begin{bmatrix}
\cos\!\big(\mu(d)(L_{n'}\!-\!L_n)\big)
&
j\sin\!\big(\mu(d)(L_{n'}\!-\!L_n)\big)
\\[0.2em]
j\sin\!\big(\mu(d)(L_{n'}\!-\!L_n)\big)
&
\cos\!\big(\mu(d)(L_{n'}\!-\!L_n)\big)
\end{bmatrix}.
\end{align}
Substituting the above into the dominant coherent term in
\eqref{eq:self_cross_coupling_gain} gives
\begin{align}
&2\Re\!\left\{
\bm h_{k,n}^{\mathsf H}
e^{j\bm C(L_{n'}-L_n)}
\bm h_{k,n'}
\right\}
\nonumber\\
=&\;
2\cos\!\big(\mu(d)(L_{n'}-L_n)\big)
\Re\!\left\{
\bm h_{k,n}^{\mathsf H}\bm h_{k,n'}
\right\}
\nonumber\\
&\;
-2\sin\!\big(\mu(d)(L_{n'}-L_n)\big)
\Im\!\left\{
\bm h_{k,n}^{\mathsf H}
\begin{bmatrix}
0&1\\
1&0
\end{bmatrix}
\bm h_{k,n'}
\right\}.
\end{align}

Next, using \eqref{eq:effective_channel_section}, and evaluating the
free-space propagation distances at a fixed reference geometry, the
phase difference between the section-$n$ and section-$n'$ responses is
approximated by the average propagation-phase difference across the
two waveguides, namely $\frac{1}{2}\sum_{m=1}^{2}
\Big(
k_0(r_{k,m_n}-r_{k,m_{n'}})
+
\beta_g(x_{m_n}^{(\mathrm{PA})}-x_{m_{n'}}^{(\mathrm{PA})})
\Big).$
Hence, the dominant coherent term can be approximated as
\begin{align*}
&2\Re\!\left\{
\bm h_{k,n}^{\mathsf H}
e^{j\bm C(L_{n'}-L_n)}
\bm h_{k,n'}
\right\}
\nonumber\\
\approx&
\;
2\|\bm h_{k,n}\|\,\|\bm h_{k,n'}\|
\cos\Bigg[
\mu(d)(L_{n'}-L_n)
\nonumber\\
&
-\frac{1}{2}\sum_{m=1}^{2}
\Big(
k_0(r_{k,m_n}-r_{k,m_{n'}})
+
\beta_g(x_{m_n}^{(\mathrm{PA})}-x_{m_{n'}}^{(\mathrm{PA})})
\Big)
\Bigg].
\end{align*}

Therefore, this coherent contribution is maximized when the cosine
term equals one, i.e.,
\begin{align}
\mu(d)(L_{n'}-L_n)
&=
\frac{1}{2}\sum_{m=1}^{2}
\Big[
k_0(r_{k,m_n}-r_{k,m_{n'}})\nonumber \\
&+
\beta_g(x_{m_n}^{(\mathrm{PA})}-x_{m_{n'}}^{(\mathrm{PA})})
\Big]
+2\pi p,
\end{align}
where $p\in\mathbb Z$.
Finally, by using $\mu(d)=\mu(a)e^{-\alpha_y(d-a)},$
we obtain
\begin{align}
e^{-\alpha_y(d-a)}
=&\frac{1}{\mu(a)(L_{n'}-L_n)}
\Bigg[
\frac12\sum_{m=1}^{2}
\Big(
k_0(r_{k,m_n}-r_{k,m_{n'}})
\nonumber\\
&+\beta_g(x_{m_n}^{(\mathrm{PA})}
-x_{m_{n'}}^{(\mathrm{PA})})
\Big)+2\pi p
\Bigg].
\end{align}

Taking the natural logarithm on both sides and rearranging yields
\eqref{eq:closed_form_waveguide_spacing}.
\end{proof}

\begin{proof}[Proof of Lemma~\ref{lem:coupling_saturation}]
For the equally spaced deployment, the coupling
matrix satisfies $[\bm C]_{m,m}=0$ and
$[\bm C]_{m,m'}=\mu(d)e^{-\alpha_y(|m-m'|-1)d}$
for $m\neq m'$.
Since $\bm C$ is real symmetric and entrywise
nonnegative, its spectral norm satisfies
$\|\bm C\|_2=\lambda_{\max}(\bm C)$.
Evaluating the Rayleigh quotient at the
normalized all-ones vector
$\bm 1_M/\sqrt M$ gives
\begin{align}
\|\bm C\|_2
\ge&
\frac{2\mu(d)}{M}
\sum_{\ell=1}^{M-1}
(M-\ell)e^{-\alpha_y(\ell-1)d}
\nonumber\\
=&\frac{2\mu(d)}{1-e^{-\alpha_y d}}
\left[
1-\frac{1-e^{-\alpha_y Md}}
{M(1-e^{-\alpha_y d})}
\right],
\end{align}
where the last equality follows from
the finite geometric series.

For the upper bound, the symmetry of $\bm C$
implies $\|\bm C\|_2\leq\|\bm C\|_\infty$.
The absolute row sum of the $m$-th row is
\begin{align*}
&\sum_{m'=1}^{M}|[\bm C]_{m,m'}|
\!=\!\mu(d)\!\left[
\sum_{\ell=1}^{m-1}e^{-\alpha_y(\ell-1)d}
\!+\!\sum_{\ell=1}^{M-m}e^{-\alpha_y(\ell-1)d}
\right]
\nonumber\\
=&\frac{\mu(d)}{1-e^{-\alpha_y d}}
\left[
2-e^{-\alpha_y(m-1)d}
-e^{-\alpha_y(M-m)d}
\right].
\end{align*}
This row sum increases toward the center
of the matrix and decreases thereafter.
Hence, its maximum is attained at
$m=\lfloor(M+1)/2\rfloor$ or
$m=\lceil(M+1)/2\rceil$, yielding
\begin{align*}
\|\bm C\|_2
\leq&\|\bm C\|_\infty
\nonumber\\
=&\frac{\mu(d)}{1-e^{-\alpha_y d}}
\Bigg[
2-e^{-\alpha_y d\left\lfloor
\frac{M-1}{2}\right\rfloor}
-e^{-\alpha_y d\left\lceil
\frac{M-1}{2}\right\rceil}
\Bigg].
\end{align*}
Combining the two bounds completes the proof.
\end{proof}

\begin{proof}[Proof of Lemma~\ref{lem:rate_bound_scaling}]
With the transverse origin at the array center,
the equally spaced waveguides are located at
$y_m^{(\mathrm{WG})}=(m-(M+1)/2)d$.
For $\zeta=3$, the geometric sum in
\eqref{eq:rate_geometry_upper_bound}
can be approximated by midpoint integration:
\begin{align}
&\sum_{m=1}^{M}
\left[
1+\left(
\frac{y_m^{(\mathrm{WG})}-y^{(\mathrm{AE})}}
{D_z}
\right)^2
\right]^{-5/2}
\nonumber\\
\approx&
\frac{1}{d}
\int_{-Md/2}^{Md/2}
\left[
1+\left(
\frac{y-y^{(\mathrm{AE})}}{D_z}
\right)^2
\right]^{-5/2}
\mathrm dy.
\label{eq:rate_scaling_integral}
\end{align}
Using the integral identity
$\int(1+u^2)^{-5/2}\mathrm du
=\mathcal F(u)+\mathrm{const}$,
we obtain
\begin{align}
&\frac{1}{d}
\int_{-Md/2}^{Md/2}
\left[
1+\left(
\frac{y-y^{(\mathrm{AE})}}{D_z}
\right)^2
\right]^{-5/2}
\mathrm dy
\\
=&\frac{D_z}{d}
\Bigg[
\mathcal F\left(
\frac{Md/2-y^{(\mathrm{AE})}}{D_z}
\right)
+
\mathcal F\left(
\frac{Md/2+y^{(\mathrm{AE})}}{D_z}
\right)
\Bigg].\nonumber
\end{align}
Substituting this result into
\eqref{eq:rate_geometry_upper_bound}
yields \eqref{eq:rate_bound_scaling_M}.
For $Md\ll D_z$, expanding the integrand
in \eqref{eq:rate_scaling_integral} around
the array center and retaining the
leading-order term gives
\begin{align*}
&\frac{1}{d}
\int_{-Md/2}^{Md/2}
\left[
1+\left(
\frac{y-y^{(\mathrm{AE})}}{D_z}
\right)^2
\right]^{-5/2}
\!\!\!\!\!\!\mathrm dy
\approx
\frac{M}{
\left[
1+\left(
\frac{y^{(\mathrm{AE})}}{D_z}
\right)^2
\right]^{5/2}
}.
\end{align*}

For $Md/2-|y^{(\mathrm{AE})}|\gg D_z$,
the agent is sufficiently far from both
array edges, allowing the integration limits
to be extended to infinity. Hence,
\begin{align}
&\frac{1}{d}
\int_{-Md/2}^{Md/2}
\left[
1+\left(
\frac{y-y^{(\mathrm{AE})}}{D_z}
\right)^2
\right]^{-5/2}
\mathrm dy
\nonumber\\
\approx&
\frac{D_z}{d}
\int_{-\infty}^{\infty}
(1+u^2)^{-5/2}\mathrm du
=\frac{4D_z}{3d}.
\end{align}
Substituting the two asymptotic expressions
into \eqref{eq:rate_geometry_upper_bound}
yields \eqref{eq:rate_bound_asymptotic}.
\end{proof}

\begin{proof}[Proof of Corollary~\ref{cor:waveguide_number_range}]
For each physical agent $k$, the minimum-rate
requirement $R_k\ge R_0$ implies
\begin{align}
P|\bm h_k^{\mathsf H}\bm w_k|^2
\ge&
(2^{R_0}-1)
\Bigg(
P\sum_{j\neq k}
|\bm h_k^{\mathsf H}\bm w_j|^2
+\sigma^2
\Bigg)
\nonumber\\
\ge&
(2^{R_0}-1)\sigma^2.
\end{align}
Applying the Cauchy--Schwarz inequality
$|\bm h_k^{\mathsf H}\bm w_k|^2
\le\|\bm h_k\|^2\|\bm w_k\|^2$
gives
\begin{equation}
P\|\bm h_k\|^2\|\bm w_k\|^2
\ge (2^{R_0}-1)\sigma^2.
\end{equation}

From \eqref{eq:rate_geometry_upper_bound},
each term in the transverse geometric sum
is at most one, so the sum is bounded by $M$.
Consequently, the channel gain satisfies
\begin{align*}
\|\bm h_k\|^2
\le&
\frac{
M\lambda^2(\zeta+1)
e^{-\alpha_{\mathrm W}\underline{x}_1}
}{
8\pi^2D_z^2
}
\left(
1+\frac{2}{N}
\sum_{n<n'}\cos\theta_{k,nn'}
\right).
\end{align*}
Combining the above inequalities with
the total precoding power constraint
$\sum_{k=1}^{K}\|\bm w_k\|^2\le 1$
yields
\begin{align}
1\ge&
\frac{
8\pi^2\sigma^2D_z^2
(2^{R_0}-1)
e^{\alpha_{\mathrm W}\underline{x}_1}
}{
PM\lambda^2(\zeta+1)
}
\nonumber\\
&\times
\sum_{k=1}^{K}
\left(
1+\frac{2}{N}
\sum_{n<n'}\cos\theta_{k,nn'}
\right)^{-1}.
\end{align}
Rearranging and noting that $M$ is an
integer gives the lower bound in (45).

Moreover, the $M$ equally spaced waveguides
must lie within the transverse deployment
interval of width $D_y$, requiring
$(M-1)d\le D_y$.
Since $M$ is an integer, this yields
$M\le 1+\lfloor D_y/d\rfloor$.
Combining the lower and upper bounds
completes the proof.
\end{proof}

\end{document}